\documentclass{aa}   

\usepackage{natbib}
\usepackage{graphicx}
\usepackage{multirow}
\usepackage{threeparttable}
\usepackage{color}
\usepackage{pdflscape}
\usepackage{txfonts}
\begin{document}

   \title{The co-existence of hot and cold gas  in debris discs }


   \author{I. Rebollido
          \inst{1}
\and       C. Eiroa\inst{1}
\and       B. Montesinos\inst{2}
\and       J. Maldonado \inst{3}
\and       E. Villaver\inst{1}
\and       O. Absil\inst{4}
\and       A. Bayo\inst{5,6}
\and       H. Canovas\inst{1,7}
\and       A. Carmona\inst{8}
\and       Ch. Chen\inst{9}
\and       S. Ertel\inst{10}
\and       A. Garufi\inst{1}
\and       Th. Henning\inst{11}
\and       D. P. Iglesias\inst{5,6}
\and       R. Launhardt\inst{11}
\and       R. Liseau\inst{12}
\and       G. Meeus\inst{1}
\and       A. Mo\'or\inst{13}
\and       A. Mora\inst{14}
\and       J. Olofsson\inst{5,6}
\and       G. Rauw\inst{4}
\and       P. Riviere-Marichalar\inst{15}
          }
   \institute{Dpto. F\'\i sica Te\'orica, Universidad Aut\'onoma de Madrid, 
Cantoblanco,
              28049 Madrid, Spain,
              \email{isabel.rebollido@uam.es}
\and Centro de Astrobiolog\'ia (CAB, CSIC-INTA), ESAC Campus, Camino Bajo del
Castillo, s/n, 28692 Villanueva de la Ca\~nada, Madrid, Spain
\and INAF, Osservatorio Astronomico di Palermo, Piazza del Parlamento 1, 90134 Palermo, Italy
\and STAR Institute, Universit\'e de Li\`ege, F.R.S.-FNRS, 19c All\'ee du Six Ao\^ut, B-4000 Li\`ege, Belgium
\and Instituto de Física y Astronomía, Facultad de Ciencias, Universidad de 
Valparaíso,  5030 Casilla, Valparaíso, Chile
\and N\'ucleo Milenio de Formaci\'on Planetaria - NPF, Universidad de Valpara\'iso, Av. Gran Breta\~na 1111, Valpara\'iso, Chile
\and European Space Astronomy Centre (ESA), PO Box, 78, 28691 Villanueva de la Ca\~nada, Madrid, Spain
\and Universit\'e de Toulouse, UPS-OMP, IRAP, Toulouse F-31400, France
\and Space Telescope Science Institute, 3700 San Martin Drive, Baltimore, MD 21212, USA
\and Steward Observatory, Department of Astronomy, University of Arizona,  Tucson, AZ 85721, USA
\and Max-Planck-Institut für Astronomie (MPIA), Königstuhl 17, D-69117 Heidelberg, Germany
\and Department of Space, Earth and Environment, Chalmers University of Technology, Onsala Space Observatory, SE-439 92 Onsala, Sweden
\and Konkoly Observatory, Research Centre for Astronomy and Earth Sciences, 
PO Box 67, 1525 Budapest, Hungary
\and Aurora Technology B.V. for ESA, ESA-ESAC
\and Instituto de Ciencia de Materiales de Madrid (CSIC). Calle Sor Juana In\'es de la Cruz 3, E-28049 Cantoblanco, Madrid, Spain}

   \date{}

 
  \abstract
 {Debris discs  have often  been
  described as gas-poor discs as  the gas-to-dust ratio is expected to
  be  considerably lower  than  in  primordial, protoplanetary  discs.
  However,  recent  observations  have  confirmed the  presence  of  a
  non-negligible amount of  cold gas in the  circumstellar (CS) debris
  discs  around young  main-sequence  stars. This  cold  gas has  been
  suggested  to be  related  to the  outgassing  of planetesimals  and
  cometary-like objects.}
{The goal of this paper is to investigate
 the presence  of hot gas  in the  immediate surroundings of  the cold-gas-bearing debris-disc central stars.}
   {High-resolution  optical spectra of  all currently known  cold-gas-bearing
     debris-disc  systems, with  the  exception  of $\beta$  Pic  and
     Fomalhaut, have  been obtained  from La  Palma (Spain),  La Silla
     (Chile),  and  La  Luz  (Mexico)  observatories.  To  verify  the
     presence of hot gas around the  sample of stars, we have analysed
     the Ca  {\sc ii} H\&K  and the Na {\sc  i} D lines  searching for
     non-photospheric absorptions of CS origin, usually  attributed to cometary-like activity.}
{Narrow,  stable  Ca {\sc  ii}  and/or  Na {\sc  i}  absorption
  features have been detected superimposed  to the photospheric lines  
  in  10 out  of the  15 observed  cold-gas-bearing debris-disc stars.  Features  are found  at the
  radial  velocity of  the stars,  or slightly  blue- or  red-shifted, 
 and/or at the velocity of the local interstellar medium (ISM).
  Some   stars  also   present  transient   variable  events   or absorptions extended towards red wavelengths (red wings). These  are the  first detections of such  Ca {\sc  ii} features  in 7
 out  of the 15 observed  stars. 
  Although an ISM origin cannot categorically be  excluded, the results suggest
  that the stable  and variable  absorptions arise from
  relatively hot  gas located  in the CS  close-in environment  of the
  stars. 
This hot gas is detected in at least $\sim$ 80\%, of edge-on cold-gas-bearing debris discs, while in only 
$\sim$ 10\% of the discs seen close to face-on. We interpret this result as a geometrical effect, and suggest that the non-detection of hot gas absorptions in some face-on systems is due to the disc inclination and likely not to the absence of the hot-gas component. This gas  is likely  released in  physical  processes  related  in  some way  to  the  evaporation  of
  exocomets,  evaporation of  dust grains,  or grain-grain  collisions
  close to the central star.}
{}


   \maketitle
%
\section{Introduction}

Debris discs are second-generation dusty discs produced by collisions
among planetesimals leftover from planet  formation.  
They are observationally characterized by the  thermal infrared excess 
observed in main-sequence stars 
\citep[e.g.][and references therein]{backman93,wyatt08}.   
Debris discs have been described  as gas-poor discs, since in contrast
with the ``canonical'' gas-to-dust ratio 100:1 in primordial discs, that
ratio,   although  unconstrained,   should   be  significantly   lower
\citep{liseau03,matthews14}. However, following the first detection of
CO in  $\beta$ Pic \citep{vidalmadjar94,roberge00,dent14},
and in 49  Ceti \citep{zuckerman95,hughes08}, a number of   cold-gas-bearing debris-disc systems -  currently 17  objects to  our knowledge  - have  been
detected through emission  lines of O {\sc  i}, C {\sc i},  C {\sc ii}
and CO at far-infrared (FIR) and  (sub-)millimetre wavelengths, mainly thanks to Herschel,
APEX,                  JCMT                  and                  ALMA
\citep[e.g.][]{moor11,moor15a,moor17,riviere12,riviere14,roberge13,donaldson13,
cataldi14,dent14,greaves16,liemansifry16,marino16}.
We note, nevertheless, that the CO emission associated with $\eta$ Crv
requires  confirmation \citep{marino17}.  The gas in emission detected in these 17 stars is  cold, T  $\sim$
10-20 K, and  located at distances from tens to  more than 100
au from the central star. It is also remarkable that numerous optical
emission lines of Ca {\sc ii}, Na  {\sc i}, Fe {\sc i}, Cr {\sc i+ii},
Ti  {\sc   i+ii},  and  other   metals,  were  detected   towards  the
archetypical $\beta$  Pic debris  disc \citep{olofsson01,brandeker04}.
A detailed analysis of the optical  data by those authors reveals that
the optical  atomic gas seen in emission is  in Keplerian rotation and  
coexisting with the dust.

 Stars with debris discs bearing cold gas  tend  to be  young
(ages $<$50 Myr), but at least  two of them, Fomalhaut and $\eta$ Crv,
are significantly older; $\sim$ 400  and $\sim$ 1400 Myr, respectively
\citep{mamajek12,marino17}.  Most  of them,  14  out  of 17,  surround
A-type stars -  the three remaining objects are F-type  stars.  All 17
systems have high fractional  dust luminosities compared to other 
debris discs, $L_{\rm dust}/L_\star
> 10^{-4}$                 \citep[e.g.][and                 references
  therein]{kennedy14,thureau14,moor17};  and  their   spectral energy distributions (SEDs)  have  been
fitted    with   two-temperature    (warm    and   cold)    components
\citep[e.g.][]{ballering13,melis13,chen14,kennedy14,thureau14,cotten16,ballering17}.
One exception is HD 172555, where only a warm component is present, at
least      up      to      the      current      detection      limits
\citep[e.g.][]{riviere12,thureau14}.

The origin of the cold gas  is a matter of debate. In the
cases  of HD  21997 and  HD 131835  the gas  could predominantly  be a
remnant      of      the      primordial      protoplanetary      disc
\citep{kospal13,moor15a}.  However, in most cases, for example, $\beta$ Pic or
HD 181327  \citep{matra17a,marino16}, the  observed gas most likely has a
secondary  origin. In primordial discs, CO is maintained against photodissociation as long as self-shielding is high enough to maintain a substantial CO mass, but the relatively low content of gas in the cold-gas-bearing debris discs suggests that self-shielding is low, and CO should be 
photodissociated on timescales of hundreds of years \citep{vandishoeck88,roberge00,moor11},  
requiring therefore a  continuous  replenishment  to  explain the  observed  
emission in debris discs.   
A variety  of  physical processes  have  been  proposed to  explain  the
secondary origin of the  cold gas, all of them in  some way related to
the presence  of planetesimals -  vaporization of solids  in dust-dust
collisions,  photon-stimulated  desorption   of  dust,  collisions  of
volatile-rich  dust grains, and outgassing of  comet-like objects.  Those
mechanisms  could even  operate concurrently  to explain  the observed
amount of gas \citep[][and references therein]{matthews14}.

In  parallel, thanks  to  the favourable  edge-on  orientation of  the
$\beta$ Pic  disc, stable and variable  UV/optical absorption features
in   metallic   lines   have    been   known   for   several   decades
\citep{slettebak75,kondo85,hobbs85,ferlet87,lagrangehenri88,roberge00}.
The variable,  transient events,  mainly traced in  the Ca~{\sc  ii} K
line ($\lambda$\,3933.66 \AA), appear red- and, to a much less degree,
blue-shifted  with respect  to the  radial velocity  of the  star, and
might vary on timescales as short as hours. They have been interpreted
as due to the gas released  by the evaporation of exocomets grazing or
falling onto the  star, {\em the Falling  Evaporating Bodies} scenario
or FEBs \citep{ferlet87,beust90,kiefer15}; Those  FEBs would have been
driven into  the vicinity of  the star by  the perturbing action  of a
larger  body,  a  planet  \citep{beust91,beustmorbidelli00}.   Similar
transient absorption  events have been  observed towards more  than 20
A-type   stars    \citep[e.g.][Welsh   \&   Montgomery,    MNRAS,   in
  press]{redfield07,  roberge08,kiefer14,welsh15,eiroa16}. Related  to
this, evidence of exocomet transits  based on Kepler light curves of
the  F2 V  stars  KIC  3542116 and  KIC  11083727  have recently  been
presented by \cite{rappaport17}.

With respect to  the stable absorption in $\beta$ Pic,  in addition to
the Ca {\sc ii}  H\&K lines at the core of  the photospheric lines that
share the  radial velocity  of the  star, similar  absorption features
appear in  UV/optical lines of  elements such as C,  O, Na, Fe,  and many
other                         metallic                         species
\citep{vidalmadjar94,vidalmadjar17,lagrange98,brandeker04,roberge06}.
This  gas would  be  located  at distances of  $\sim$  0.5  au, would  be
relatively   hot  with   temperatures  of   $\sim$  1000--2000   K
\citep{hobbs88,beust98,vidalmadjar17},  and  could   be  generated  by
evaporation  of  comet-like   bodies  or  dust  grains   in  the  disc
\citep{fernandez06}.  However,  this hot  gas should rapidly  be blown
away by the  strong radiation pressure from the star  unless a braking
mechanism is at work.  \cite{fernandez06} have suggested a solution to
this problem, showing that the  gas affected by the radiation pressure
is  largely ionized  and  couples  into a  single  ionic  fluid. As  a
consequence, the radiation pressure is  reduced and the fluid could be
self-braking if  it has enhanced  carbon abundance.  Such  a scenario is
supported  by  the   observed  overabundance  of  C   in  $\beta$  Pic
\citep{roberge06, brandeker11}. A similar mechanism could be acting in
the case of 49 Ceti \citep{roberge14}.

This work  presents high-resolution  optical spectra of  the currently
known cold-gas-bearing debris-disc  stars. Our  aim is to  search for
signposts of  hot gas,  which would suggest  the presence  of close-in
cometary-like  material  around those  stars.  Thus,  the analysis  is
mainly centered on  the Ca {\sc ii} K ($\lambda$  3933.66 \AA) line as
it  is the  most  sensitive one  to trace  both  transient and  stable
absorptions, but  we also refer  to the Ca {\sc  ii} H ($\lambda$
3968.47 \AA) and  Na {\sc i} D  ($\lambda\lambda$ 5889.95/5895.92 \AA)
lines. The spectra have been obtained  in the working frame of a large
survey  aiming at  detecting and  monitoring cometary-like  absorption
events in  a sample of  main-sequence stars,  most of them  A-type but
also some FG-type (Rebollido et al., in preparation).

\section{Observations}

\begin{table}[ht!]
{\small
\caption{Log of observations. Number of spectra per instrument and signal-to-noise 
ratio (S/N) in brackets of the median spectra.}
\label{table:log}
{\centering
\begin{tabular}{ll}
\hline
\hline
Star      & Number of Spectra (S/N)\\
\hline
HD 9672   & 31 H (296) + 7 FI (161)  \\
HD 21997  & 13 H (133) + 3 FE (281) + 9 HE (130)\\
HD 32297  & 15 H (98), 4 FI (97) \\
HD 109085 & 3 H (85)\\
HD 110058 & 8 FE (280)\\
HD 121191 & 1 FE (90)\\
HD 121617 & 2 FE (103)\\
HD 131488 & 2 FE (110)\\
HD 131835 & 13 FE (390)\\
HD 138813 & 20 H (205) + 13 FE (447) + 4 FI (102)  \\
HD 146897 & 2 H (7) + 4 FE (112)\\
HD 156623 & 13 FE (314)\\
HD 172555 & 17 FE (380)\\
HD 181296 & 16 FE (373)\\
HD 181327 & 5 FE (241)\\
\hline
\end{tabular}
}
\tablefoot{
H: HERMES; FE: FEROS; FI: FIES; HE: HEROS.}\\
}
\end{table}

\begin{table*}[ht!]
\caption{Stellar properties of the observed cold-gas-bearing debris-disc stars}
\label{tbl:parameters}
\centering
\begin{tabular}{lllllllllll}
\hline
\hline
Star      & Other Name & SpT    & d    & $V$  &$B$   & T$_{\rm eff}$&$\log g$& v$\sin i$  & Age$^a$ & Cold gas detection $^b$\\
          &            &        & (pc) &(mag.)&(mag.)& (K)        &        & (km/s)     & (Myr) &  \\
\hline                                                                                   
HD 9672   & 49 Ceti    & A1V    & 59.4 & 5.61 & 5.67 & 9110   & 4.25   & 186        & 40 (1) &   (1) CO   \\
HD 21997  & HIP 16449  & A3IV/V & 68.3 & 6.37 & 6.50 & 8610   & 4.10   &  60        & 42 (2) &   (2) CO   \\
HD 32297  & HIP 23451  & A0V    &136.2 & 8.14 & 8.32 & 7910   & 4.05   &  90        & < 30 (3) & (3) CO \\
HD 109085 &$\eta$ Crv  & F2V    & 18.3 & 4.31 & 4.69 & 6950   & 4.20   &  60        & 1400 (4) & (4) CO \\
HD 110058 & HIP 61782  & A0V    &188.7 & 7.97 & 8.12 & 9720   & 4.45   & 150        & 15 (5)  &  (5) CO \\
HD 121191 & SAO 241295 & A5IV/V &135.9 & 8.16 & 8.40 & 7970   & 4.25   &  65        & 16 (5) &   (6) CO \\
HD 121617 & SAO 224570 & A1V    &128.2 & 7.29 & 7.36 & 9285   & 4.45   &  90        & 16 (5)  &  (6) CO \\
HD 131488 & SAO 320038 & A1V    &150.0 & 8.00 & 8.09 & 9050   & 4.21   & 120        & 16 (5) & (6) CO \\
HD 131835 & HIP 73145  & A2IV   &145.6 & 7.86 & 8.05 & 8610   & 4.35   & 105        & 16 (5)  &   (7) CO         \\
HD 138813 & HIP 76310  & A0V    &134.4 & 7.30 & 7.37 & 9685   & 4.45   & 130        & 10 (5)  &   (5) CO  \\
HD 146897 & HIP 79977  & F2/3V  &122.4 & 9.11 & 9.58 & 6700   & 4.30   &  55        & 15 (6)  &   (5) CO  \\
HD 156623 & HIP 84881  & A0V    &118.3 & 7.26 & 7.35 & 9580   & 4.10   &  80        & 16 (5)  &  (5) CO  \\
HD 172555 & HIP 92024  & A7V    & 28.6 & 4.77 & 4.97 & 7900   & 4.05   & 120        & 23 (7)  &   (8) OI  \\
HD 181296$^c$&$\eta$ Tel& A0V+M7& 48.2 & 5.02 & 5.04 & 10400  & 4.30   & 230   & 23 (7) &        (9) CII \\
HD 181327 & HIP 95270  & F6V    & 48.6 & 7.04 & 7.50 & 6650   & 4.40   &  28        & 23 (7)  &  (10) CO \\
\hline
\end{tabular}

\tablefoot{$^{(a)}$ References for ages: (1) \cite{torres08}; (2) \cite{bell15} ; (3) \cite{kalas05}; (4) \cite{marino17}; (5) \cite{pecaut16}; (6) \cite{pecaut12}; (7) \cite{mamajek14}.
$^{(b)}$ References for cold gas detection: (1) \cite{zuckerman95}; (2) \cite{moor11}; (3) \cite{greaves16}; (4) \cite{marino17}; (5) \cite{liemansifry16}; (6) \cite{moor17}; (7) \cite{moor15a}; (8) \cite{riviere12}; (9) \cite{riviere14}; (10) \cite{marino16}   
$^{(c)}$ The M7  brown dwarf companion of $\eta$ Tel is located at an angular 
distance of $\sim 4\arcsec$ \citep{neuhauser11}.}
\end{table*}

High-resolution  optical spectra of 15  out of the currently  known 17 cold-gas-bearing debris-disc stars were obtained with  several fibre-fed high-resolution   spectrographs  and   telescopes   located  at   different
observatories  (Table \ref{table:log})  .   Namely, the  spectrographs
HERMES at Mercator and FIES at  NOT (Roque de Los Muchachos, La Palma,
Spain), FEROS  at MPG/ESO 2.2  m (La Silla,  Chile), and HEROS  at the
robotic telescope TIGRE (La Luz,  Mexico).  A series of observing runs
were carried out at each telescope: Mercator (2015 September/December;
2016 January/March/July;  2017 March/April), NOT  (2016 January/July);
MPG/ESO  2.2 m  (2015 October;  2016 March;  2017 April);  TIGRE (2015
September/November).  The number of spectra  per star and telescope is
given in Table \ref{table:log}; a  more detailed log with the specific
dates  (UT) of  the observations  of each  star and  the corresponding
instrument is given in Table B.1 of Appendix B.  Spectral range
and resolution  of the  different spectrographs are:  HERMES, spectral
resolution     R$\sim$85000     covering    the     range     $\lambda
\lambda\sim$370--900 nm  ~\citep{hermes}; FIES,  R$\sim$67000 covering
$\lambda \lambda  \sim$370--830 nm ~\citep{fies};  FEROS, R$\sim$48000
and  $\lambda   \lambda\sim$350--929  nm  \citep{feros};   and  HEROS,
R$\sim$20000  and $\lambda  \lambda  \sim$350--880 nm  ~\citep{tigre}.
The spectra  have been  reduced using the  available pipelines  of the
corresponding  instrument; they  include the  usual steps  for echelle
spectra, such  as  bias  substraction,  flat-field  correction,  cosmic  ray
removal, and order extraction.   Wavelength calibration is carried out
by means of Th-Ar lamp spectra taken  at the beginning and end of each
night.  Finally, barycentric  corrections have been applied  to all of
them.  Telluric lines  heavily affect the wavelength  range around the
Na\,{\sc i}  D lines. To correct  for them we have  taken advantage of
the fact that the spectra were obtained at many different epochs
of the year.  Thus, it is  easy to identify and remove telluric lines,
as they shift in wavelength when the barycentric correction is applied
but the  stellar Na  {\sc i}  D lines remain  at the  same wavelength;
furthermore,  both lines  of the  Na {\sc  i} doublet  share the  same
radial velocity. Table \ref{table:log}  also gives the signal-to-noise
ratio (S/N)  of the  median  spectra  achieved for  each  instrument in  two
``continua'' at each side of the Ca {\sc ii} K line. Those values, and
those of the individual spectra,  are mostly affected by the observing
weather conditions.

\section{Stellar properties of the observed sample}

Table  \ref{tbl:parameters}  shows  some   of  the  observational  and
fundamental properties of the sample  of stars presented in this work.
Names, spectral  types, and $B$  and $V$ magnitudes  have been  taken from
SIMBAD \citep{wenger00},  while the  distances are estimated  from the
revised Hipparcos parallaxes \citep{leeuwen07} and/or the Gaia archive
\citep{lindegren16},    except    for    HD    131488    taken    from
\cite{melis13}. We note that there  is a large discrepancy between the
Gaia distance of HD 110058 of 188.7  pc and the Hipparcos one of 107.4
pc. Ages from the literature and the corresponding references 
are also given in Table \ref{tbl:parameters}, as well as the list
  of  papers reporting the cold  gas detection.
   
 $T_{\rm eff}$ , $\log g,$ and $v \sin i$ have been estimated using the
spectral  synthesis  programs {\sc  atlas}  and  {\sc synthe}  (Kurucz
1993),  fed  with the  models  describing  the stratification  of  the
stellar atmospheres (Castelli  \& Kurucz 2003). A grid  of models with
temperatures from 6000  to 10500 K, step  100 K, and log  g= 3.5, 4.0,
4.5 was created and prepared to be broadened with any value of $v \sin
i$.  Solar abundances were used for all objects, with the exception of
HD  32297  and  HD  146897  as the  fits  clearly  pointed  towards  a
metallicity different  from solar.  For these two  stars, metallicities
[Fe/H] $-0.5$ and  $-0.2$ were used, respectively (Gebran  et al. 2016
give a value $-0.7$ for HD 32297).

For stars hotter  than $\sim\!8000$ K, the first step  was to find the
three models whose temperatures match the depth of the photospheric Ca
{\sc ii}  K line  for the  three values  of $\log  g$.  The  widths of
H$\gamma$, H$\delta$, H[8-2]  and H[9-2] were measured  in the stellar
spectra and in the models. Values  of stellar gravities were found for
each  line by  interpolating the  width of  the stellar  lines in  the
models' measurements. The final value of the gravity, $\log g$, is the
average  of   the  individual   determinations  for  each   line,  the
uncertainty in the  gravity being the standard error of  the mean. The
value of  $\log g\pm\Delta\log  g$ is then  interpolated in  the three
models  mentioned  above   to  obtain  the  best   fit  value  $T_{\rm
  eff}\pm\Delta T_{\rm  eff}$. The uncertainties  in $v \sin  i$ have
been estimated  by comparing the width  of the Mg {\sc  ii} 4481 \AA{}
absorption in the stellar spectrum with that of models computed with a
range of rotation rates so as to allow the relative differences between
star and model to be the order of 5\%.

For the F  stars, the approach, being  qualitatively similar, accounts
for  the different  dependence of  the calcium  and hydrogen  lines on
temperature and gravity as compared  with late-B or A-type stars. Once
an acceptable model fitting the Ca  {\sc ii} HK and the hydrogen lines
is  found, uncertainties  are estimated  by changing  model parameters
until the departures from the best  fit are $\sim\!10$\%. The value of
$v \sin i$ is estimated for this kind of stars by comparing the widths
of  weak photospheric  lines  with  those of  models  and following  a
similar  criterion  as for  hotter  stars.  Final typical  errors  are
$\sim\!$100 K,  $\sim\!$0.20, and $\sim\!5$  km/s for $T_{\rm  eff}$ ,
$\log g$ and $v \sin i$, respectively.

\section{Results}

\setcounter{figure}{0}
\begin{figure*}[ht] 
\centering
\scalebox{0.75}{\includegraphics{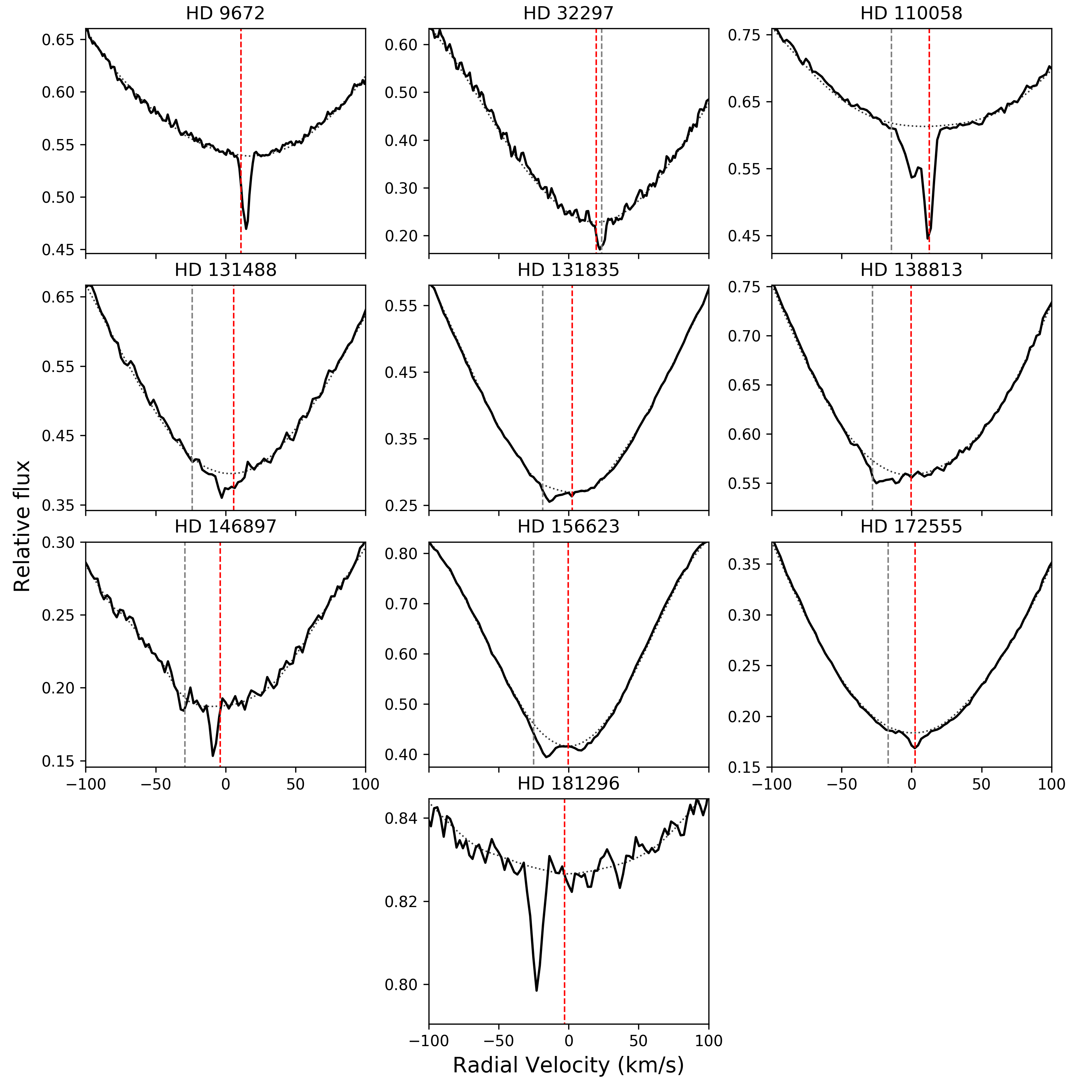}}
\caption{Median Ca {\sc ii} K  line profiles of the gas-bearing debris-disc  stars  with  non-photospheric absorption  (black 
    solid line) together with a  Kurucz photospheric model (black dotted
  line)    with     stellar    parameters    according     to    Table
  \ref{tbl:parameters}.  Dashed   lines  in  all  panels   indicate  the
  corresponding radial velocities of the stars (red), and the projected
  velocities of  the local  ISM (grey) after  the Colorado model \citep{redfield08}. Velocities are in the 
  reference frame of the Ca {\sc ii} K  line. v$_{\rm  rad}$ and v$_{\rm ISM}$ 
  are coincident for HD 9672 (see Table \ref{tbl:velocities}). HD 181296 has 
  no ISM lines in the line of sight (see Table 3 and Sect 5.2 for a discussion). The
  same convention for v$_{\rm  rad}$ and v$_{\rm ISM}$ is used  in all further figures.}
\label{fig:1} 
\end{figure*}

\begin{table*}[!ht]
\caption{Radial  velocities and equivalent widths (EWs)  of  the  Ca {\sc  ii}  H\&K and Na {\sc i} D non-photospheric
  absorptions, radial velocities of the stars, and projected velocities
  of ISM clouds in the Colorado model. The FWHM velocity dispersion 
of the Ca {\sc  ii} K lines is also given. }
\label{tbl:velocities}
\centering
\begin{tabular}{lrcrlclr}
\hline
\hline
\noalign{\smallskip}
\multicolumn{7}{c}{Stars with non-photospheric  absorptions} \\
\noalign{\smallskip}
\hline
Star          &v$_{{\rm Ca {\sc II} K}}$&EW(K/H) & $\Delta$v $_{{\rm Ca {\sc II} K}}$ &v$_{{\rm Na {\sc I} D}}$&EW(D2/D1) $^c$  &v$_{\rm rad}$   &v$_{\rm ISM}$ $^d$ \\
              & (km/s)             & m\AA    &  (km/s) &    (km/s)     &   m\AA     & (km/s)       & (km/s)       \\
\hline                                                                               
HD 9672$^a$  &13.5$\pm$2.3               & 10.6/6.0     &7.4    & --               & <1.5     & 11.0$\pm$2.0  & 11.01 (LIC) \\
HD 32297$^a$ &22.7$\pm$3.0              & 19.4/8.9      &9.7    & 23.8$\pm$2.5         &93.0/84.0  & 19.7$\pm$2.0  & 23.59 (LIC)  \\
HD 110058&12.7$\pm$3.3              & 27.0/24.1 &6.8   & 12.5$\pm$2.8           &46.0/39.1      & 12.6$\pm$2.4  & --14.46 (G)  \\
             &1.2$\pm$3.3               & 19.9/10.0     &12.6   & 0.9$\pm$2.8            &34.0/18.9     &               &              \\
HD 131488&4.0$\pm$2.7               &5.7/--     &9.6   & 5.3$\pm$2.2             &42.1/34.2      &  5.8$\pm$1.5  & --23.91 (G)  \\
             &-3.2$\pm$2.7               &4.0/--        &4.2   & --               & <1.6     &               &              \\
HD 131835&2.2$\pm$2.7              &0.7/--      &2.3   & --               & <0.9     & 2.6$\pm$1.4  &--18.56 (GEM)  \\
             &--5.9$\pm$2.7           &2.1/1.4  &7.9  & --               & <0.9     &               &             \\
             &--13.8$\pm$2.7          &8.3/4.1  &7.4  & --11.8$\pm$2.0         &41.0/32.1      &               &             \\
HD 138813    &--27.5$\pm$3.0            &5.6/-- &18.7   &--12.0$\pm$2.5          &18.0/15.4  &--0.4$\pm$2.0  &--27.82 (G)   \\
HD 146897    &--8.0$\pm$2.7            &12.4/-- &7.2  &--8.8$\pm$2.1           &118.2/80.0     &--4.0$\pm$1.5  &--29.19 (G)   \\
             &--30.0 $\pm$2.7            &5.9/--        &4.6   &  --              &   <0.8   &               &             \\
HD 156623$^a$&--15.0$\pm$2.7             &9.0/3.7       &18.7   &  --              &   <0.7   &--0.2$\pm$1.5  &--24.89 (G)   \\
             &8.1 $\pm$2.7              &4.3/-- &6.0 $^e$   &  --              &   <0.7   &               &             \\
HD 172555$^a$& 2.3$\pm$3.0             &9.9/4.0 &9.7  &  --              &   <0.7   &  2.6$\pm$1.9  &--16.84 (G)   \\
HD 181296$^b$&--22.8$\pm$3.8            &3.9/2.2        &8.4  &  --              &   <0.7   &--3.0$\pm$3.0  &--18.51 (G)   \\
\hline
\noalign{\smallskip}
\multicolumn{7}{c}{Stars without non-photospheric absorptions} \\
\noalign{\smallskip}
\hline
HD 21997     &--                 &--       &--     &--           &--         & 17.3$\pm$2.0  & 15.87 (LIC)\\
HD 109085    &--                 &--       &--      &--          &--         &--0.4$\pm$1.0  &--0.11 (GEM)\\ 
HD 121191    &--                 &--       &--      &--          &--         & 12.0$\pm$1.4  & --18.34 (G)\\
HD 121617    &--                 &--       &--      &--          &--         &  7.8$\pm$1.6  & --19.84 (G)\\
HD 181327$^b$&--                 &--       &--      &--          &--         &  0.2$\pm$1.5  &--18.51 (G) \\
\hline
\end{tabular}
\tablefoot{$^{(a)}$ Stars with FEBs.$^{(b)}$ The line of sight towards these two 
stars does not traverse any ISM cloud in the Colorado model but it passes within 20$\degr$ of 
several clouds.  Among these, cloud G is the closest one in velocity to these stars. 
$^{(c)}$ Upper limits are shown for those stars without Na non-photospheric absorption. 
$^{(d)}$ Names within parentheses are ISM clouds in the Colorado model \citep{redfield08}.
{$^{(e)}$}The velocity dispersion for this absorption is highly variable, and the value shown 
here is the value obtained from the median spectra.}
\end{table*}

Narrow, stable, non-photospheric absorption features at the centre of,
or  slightly  blue-/red-shifted  with  respect  to,  the  photospheric
Ca~{\sc ii} K absorption  line have been detected in 10  out of the 15
observed  debris-disc  systems  (Table   \ref{tbl:velocities}).  The
non-photospheric absorption profile of some stars presents a red wing,
suggesting  that  the  observed  profile is  a  blend  of  independent
absorptions.   In  addition, a  few  stars  show transient  absorption
events.  In  the following we describe  the results of the  stars with
and without non-photospheric absorptions separately.
\subsection{Stars with non-photospheric Ca~{\sc ii} absorptions}

Figure~\ref{fig:1} shows the observed Ca~{\sc ii} K line profiles of the
observed stars  with narrow, stable central  absorptions, superimposed
to  the  rotationally  broadened   photospheric  line.   The  plotted,
observed profile is the median of the spectra of each star.  As far as
we know,  this is  the first  time that the  Ca~{\sc ii}  K absorption
feature  is   reported  towards  HD\,32297,   HD\,131488,  HD\,131835,
HD\,138813,   HD\,146897,    HD\,156623,   and    HD\,181296   ($\eta$
Tel). 
Welsh \& Montgomery (MNRAS, in  press) have also recently detected the
narrow  absorption towards  $\eta$  Tel.   Previous narrow  absorption
detections have  been published in  the cases of 49\,Ceti,  HD 110058,
and    HD    172555   \citep{montgomery12,kiefer14,hales17}.     Table
\ref{tbl:velocities} gives equivalent widths (EWs), radial velocities,
and velocity dispersion  (full width at half maximum (FWHM)) of all non-photospheric  Ca~{\sc ii} K
features.  The table also gives EWs of the Ca~{\sc ii} H line in those
cases where the line can  be distinguished from the strong H$\epsilon$
Balmer line, as well  as EWs and velocities of the Na  {\sc i} D lines
when  detected. In  the case  of non-detection in  Na {\sc  i}, upper
limits  of  the D2  line  (the  strongest  one  of the  doublet)  were
calculated  as EW  of  Gaussians with  parameters
  determined by the instrumental resolution (width) and the 3$\sigma$
rms noise in the wavelength region  (depth). In case of detection, EWs
are the average of independent  measurements made with three different
software packages -  MIDAS, dipso and Gaussian fits  made with Python.
Estimated uncertainties are of the  order of 10\%. Velocity dispersions
in Table  \ref{tbl:velocities}  were calculated  in the  median
spectra as the FWHM  of fitted Gaussians.  Radial
    velocities   were   measured  as   the   central   value  of   the
    non-photospheric absorption.  Uncertainties  were estimated as the
    square sum of the uncertainties in the radial velocity of the star (see below)
    and the size of the pixel in velocity.

Stable  absorptions might  originate in  CS discs  sharing the  radial
velocity  of   the  central  star,   or  in  the  complex   local  ISM
\citep[e.g.][]{redfield08}.  In  order to try to  decipher which of these alternatives corresponds to reality, radial velocities of the stars were estimated using
the Balmer  lines and the velocity  shift of the Kurucz  models to fit
the   observed    Ca   {\sc    ii}   K   photospheric    line   (Table
\ref{tbl:velocities}).  Quoted v$_{\rm rad}$  errors correspond to the
difference between both methods. For  the sake of completeness, radial
velocities of the stars and  of the corresponding ISM velocity vectors
are also given  in Table \ref{tbl:velocities} for  those stars without
non-photospheric absorptions.   Our stellar radial  velocity estimates
are      in      good      agreement      with      previous      ones
\citep[e.g.][]{bruijne12,melis13,hales17}.    For  the   ISM  velocity
vectors,    we   have    used    Redfield    \&   Linsky's    Colorado
model\footnote{http://sredfield.web.wesleyan.edu/}  \citep{redfield08}
to estimate them  along the line of sight towards  the stars. Names of
the   corresponding   ISM  clouds   are   also   indicated  in   Table
\ref{tbl:velocities}.  In the cases of  HD 181296 and HD 181327, their
lines  of sight  do  not traverse  any cloud  in  the Colorado  model,
although they  pass within 20$\degr$ of  several clouds. 
A complementary search of ISM absorption features has been conducted in the cases of HD 131488, HD  131835, HD 138813, and  HD 146897. This search was motivated since, based on previous results and on our own spectra, the origin of the observed non-photospheric absorptions was highly ambiguous. In this context, we have analysed the  Ca {\sc ii} and Na {\sc i} lines of field stars surrounding  those four cold-gas-bearing debris discs. We note that these stars are located in the line of sight of the G or Gem Colorado clouds (Table \ref{tbl:velocities}), and are members of the Scorpius-Centaurus  OB  association  \citep{poppel10,liemansifry16}.   Details of this search are given in Appendix A.
The remaining 11 objects in the sample were excluded from this study due either to the presence of variable non-photospheric absorptions, or the unambiguous determination of a CS absorption.

\subsubsection{Individual stars}

\noindent {\em  HD 9672  (49 Ceti):}  A Ca~{\sc  ii} K  narrow, stable
absorption at the  bottom of the photospheric line  at $\sim$13.5 km/s
and  EW(Ca~{\sc ii}  K) $\sim$10.6  m\AA ~is  detected.  This  feature
coincides with the radial velocities of  the star and of the local ISM
(Fig.~\ref{fig:1}  and  Table  \ref{tbl:velocities}).  Our  result  is
similar  to  the  stable component  observed  by  \cite{montgomery12}.
Using   the   high-resolution   spectrograph    STIS   at   the   HST,
\cite{malamut14} detected  two partially blended, near-UV  Mg {\sc ii}
and Fe  {\sc ii} features with  velocities 9.0 and 14.4  km/s (Mg {\sc
  ii}), and 11.0  and 13.65 km/s (Fe {\sc  ii}), and suggested
  the second  one could be  attributed to the  CS disc.  Thus,  it is
most likely that  the absorption we detect  is a blend of  both the CS
and  the ISM  components.  The  Ca~{\sc ii}  H line  is detected  with
EW(Ca~{\sc ii})  $\sim$6.0 m\AA, which  suggests that the Ca  {\sc ii}
lines are optically thin.  None of  our spectra show any transient Ca
{\sc  ii}   absorption  in   contrast  with   the  FEBs   detected  by
\cite{montgomery12}.   Neither  stable  components  nor any  variable
features are detected in the Na {\sc i}\,D lines.
 
\noindent {\em HD 32297:} The radial  velocity of the star (19.7 km/s)
is very close  to that of the  local ISM.  The velocity and  EW of the
detected  Ca~{\sc  ii} K  stable  absorption  is $\sim$22.7  km/s  and
$\sim$19.4    m\AA,   respectively    (Fig.~\ref{fig:1}   and    Table
\ref{tbl:velocities}).   The  Ca~{\sc  ii}  H line  is  detected  with
EW(Ca~{\sc ii}\,H)\,$\sim$8.9\,m\AA.  Very  high-resolution spectra, R
$\sim$240000,  of the  Na  {\sc  i} doublet  show  two components  at
$\sim$20.5 km/s and  $\sim$24.5 km/s, attributed to the  CS medium and
the ISM,  respectively \citep{redfield07}.   This suggests  again that
the Ca~{\sc ii} absorptions  we detect are a blend of  both CS and ISM
components. \cite{redfield07}  detected variability in the  Na {\sc i}
lines; in our case, no apparent variability is detected either in the
Ca~{\sc ii} lines or in the Na {\sc i} ones.

\noindent {\em HD 110058:} Two  Ca~{\sc ii} K components at velocities
of 12.7 and 1.2 km/s are  detected with EW $\sim$27 m\AA ~and $\sim$20
m\AA, respectively (Fig.~\ref{fig:1}  and Table \ref{tbl:velocities}).
The 12.7  km/s absorption  coincides with the  radial velocity  of the
star, while  the ISM  velocity towards  this line  of sight  is -14.46
km/s.   The   same  velocity   components   have   been  reported   by
\cite{hales17} and  Iglesias et  al. (in preparation).  However, while
both  features  remain  stable  in  our spectra  (taken  in  the  same
campaign),  a comparison  with  those of  similar spectral  resolution
\citep{hales17}  suggests   that  the  strength  of   the  blueshifted
component varies.   \cite{hales17} ascribe this absorption  to the ISM
because  of the  presence  of  a similar  feature  in  the spectra  of
close-by  stars.   However, the  fact  that  its strength  might  vary
suggests that  it might well have  a CS origin. Further  monitoring of
this star  is required to  elucidate this apparent  controversy.  Both
features  are detected  in the  Ca~{\sc  ii} H  line with  \makebox{EW
  $\sim$24.0 and $\sim$10.0 m\AA,} respectively.  This result suggests
that the stronger  12.7 km/s absorption is optically  thick, while the
weaker one at 1.2  km/s is optically thin. A similar  behaviour is present
in the Na {\sc i} D lines \citep[Table \ref{tbl:velocities},][Iglesias
  et al., in preparation]{hales17}.

\noindent {\em HD 131488:} The  non-photospheric Ca~{\sc ii} K profile
shows a narrow  absorption with a red wing extending  up to $\sim$12.0
km/s,    suggesting   a    blend   of    two   independent    features
(Fig.~\ref{fig:1}).  The  Ca~{\sc ii}  K profile can  be fit  with two
Gaussians  centred at  4.0  and --3.2\,km/s,  and  EWs $\sim$5.7  and
$\sim$4.0 m\AA,  respectively (Fig. 2).   While both features  are far
away  from the  expected  ISM velocity  in the  Colorado
  model, the 4.0\,km/s feature coincides  with the radial velocity of
the   star   \citep[Table~\ref{tbl:velocities},  and][]{melis13}.    A
Ca~{\sc ii} H  absorption is indistinguishable from the  noise. The Na
{\sc i} doublet appears at a velocity  of 5.3 km/s, that is, at the radial
velocity  of  the  star  and  one of  the  Ca~{\sc  ii}  K  components
(Fig. 2). Their EWs are  \makebox{$\sim$42 m\AA ~(Na\,{\sc i} D2)} and
$\sim$\,34\,m\AA ~(Na {\sc i} D1), which  suggests that the Na {\sc i}
lines are optically  thick.  Among the  field stars around
  HD 131488 (see Appendix A), HD  132200, located at a distance of 165
  pc and at  an angular distance from HD 131488  of 1.24$\degr$, shows
  two non-photospheric  Ca~{\sc ii}  K absorptions at  velocities -6.9
  km/s  and 4.6  km/s,  which are  similar to  the  velocities of  the
  observed  features in  HD 131488.   As mentioned  before, both  stars
  belong to  the complex  Scorpius-Centaurus OB association,  and both
  show a feature at negative velocities  close to the mean velocity of
  $\sim$ --6.6 km/s that characterize the approaching face, located at
  a distance $\leq$ 76 pc, of the large expanding bubble around the OB
  association  \citep{poppel10}. Thus, the negative velocity feature of 
both stars is likely of ISM origin. At  the  same time,  the  4.6  km/s
  feature of  HD 132200 is variable,  suggesting a relation  to the
  star (see Fig. A.1, Appendix A). This fact also suggests that the 
HD 131488 4.0 km/s absorption, which coincides with the stellar radial velocity,
is likely circumstellar.

\begin{figure}[h] 
\label{fig:hd131488_feb}
\centering
\scalebox{0.6}{\includegraphics{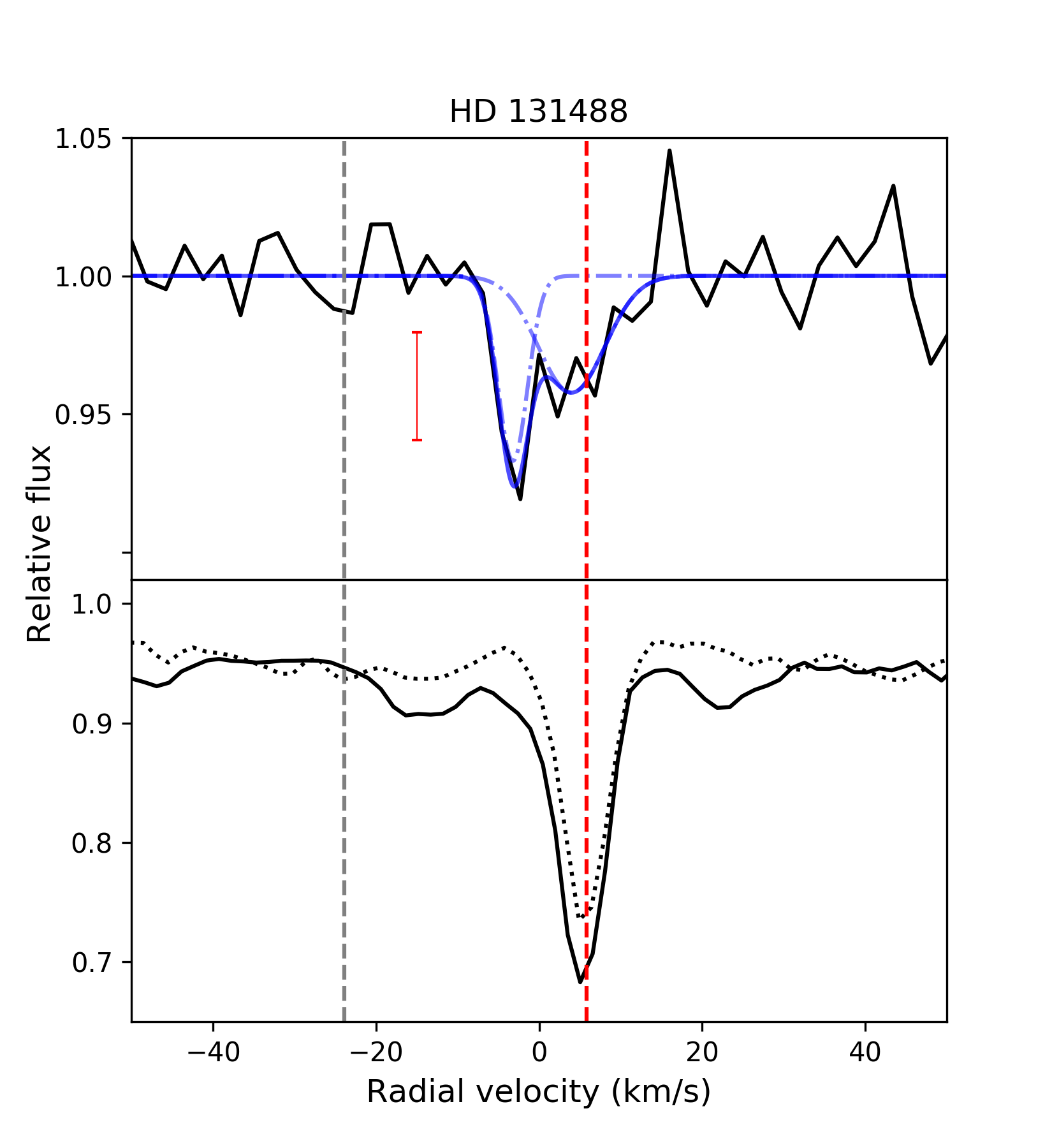}}
\caption{Top: Non-photospheric Ca {\sc ii}  K  feature profile of
  HD 131488. The observed spectrum is plotted with a black solid line. 
The Ca {\sc ii} K feature is fit with two 
Gaussians, plotted with a blue dash-dotted line; the blue continuous 
line is the sum of both Gaussians . A 3-$\sigma$
  error bar is plotted in red. Bottom:  both Na {\sc  i} D2  (continuous line)  
and Na {\sc  i} D1  (dotted line) components.}
\end{figure}

\noindent  {\em  HD  131835:}   This  star  shows  a  non-photospheric
absorption in both Ca~{\sc ii} H\&K lines with a weak red wing
extending up to $\sim$ 0.0 km/s (Figs.~\ref{fig:1} and 3).  The feature
profiles can  be fit with  two (Gaussian) components at  $\sim$ --13.8
km/s and EW $\sim$ 8.3\,m\AA, and  --5.9 km/s ~and EW $\sim$ 2.1 m\AA.
The  corresponding Ca~{\sc  ii} H  EW values  are $\sim$4.1  m\AA ~and
$\sim$ 1.4  m\AA, respectively.  A  third very weak  component, $\sim$
3$\sigma$ level, is present only in Ca~{\sc ii} K at a velocity of 2.2
km/s   and    EW   $\sim$0.7   m\AA    ~(Table   \ref{tbl:velocities},
Figs.\ref{fig:1}  and  3).   This  weak feature,  which  would  require
further  confirmation,  coincides  with  the radial  velocity  of  the
star.   The  Na   {\sc  i}   D  lines   show  a   narrow  feature   at
\makebox{$\sim$\,--11.8 km/s,}  approximately at  the velocity  of the
strongest  Ca~{\sc  ii} component  (Fig.  3).  The estimated  EWs  are
\makebox{$\sim$41 m\AA ~(Na  {\sc i} D2)} and $\sim$32  m\AA ~(Na {\sc
  i} D1), suggesting they are  partly optically thick.  We
  note that stars in the field show a non-photospheric  absorption at $\sim$
  --13 km/s (Appendix A), close to  the one observed in HD 131835, and
  that the -5.9  km/s feature is close to the  mentioned mean velocity
  of  the  approaching  face  of   the  expanding  bubble  around  the
  Scorpius-Centaurus association. Thus, both features are likely to be interstellar,
while the weak 3$\sigma$ one has an ambiguous origin (see Appendix A.)

\begin{figure}[h] 
\label{hd131835_feb}
\centering
\scalebox{0.6}{\includegraphics{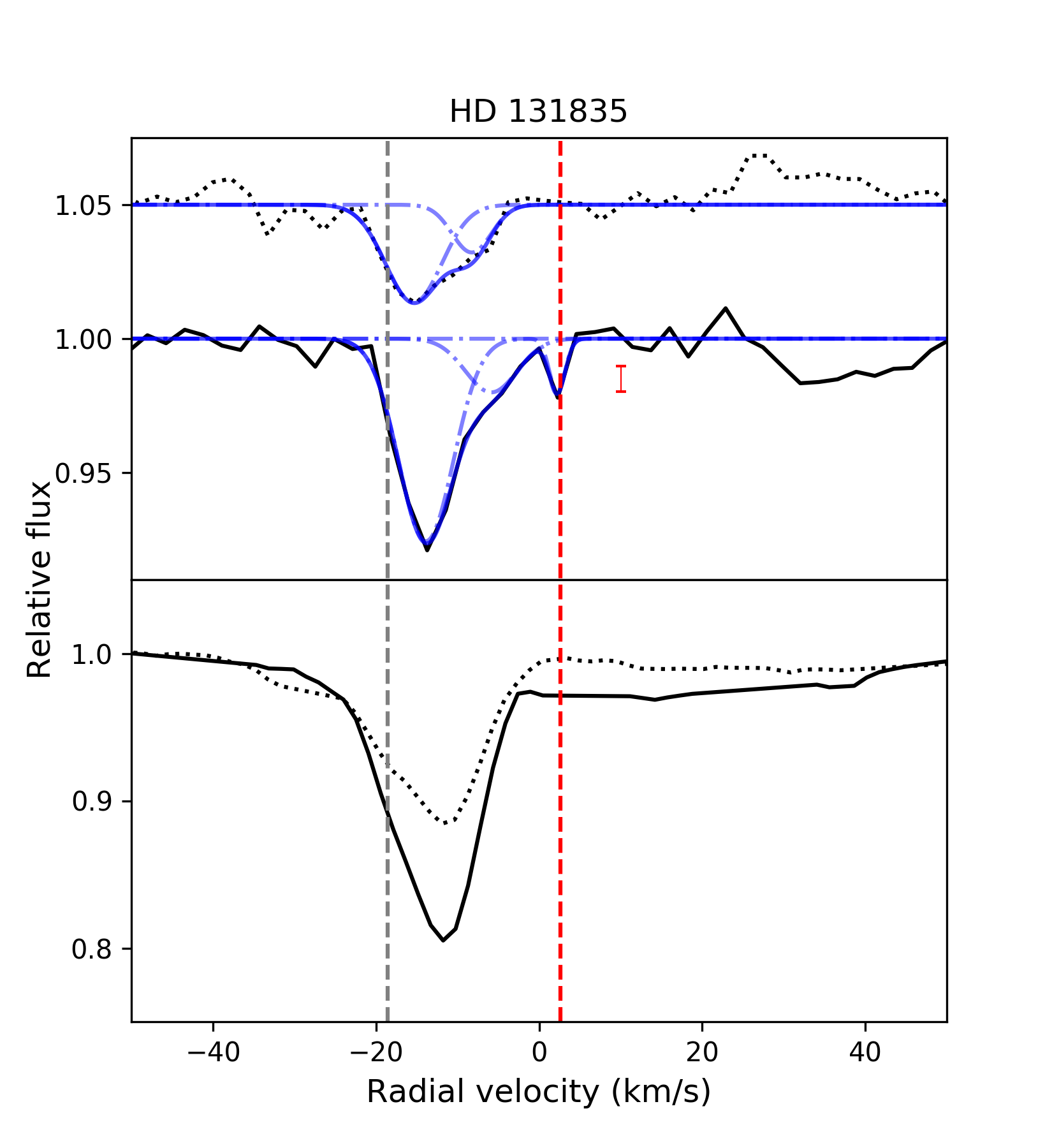}}
\caption{Top   panel:  Non-photospheric Ca   {\sc  ii}  K (black continuous line)
and H  (black dotted lined) feature profiles of   HD 131835.  Ca {\sc  ii} H has been 
displaced 0.05 relative units for clarity. Both Ca {\sc ii} features are fit with three and two 
Gaussians, respectively, plotted with  blue dash-dotted lines; the blue continuous 
lines are the sum of the Gaussians in each case. A 3-$\sigma$ error bar  is also plotted in 
red. Bottom panel: Na  {\sc i} D2
  (black continuous  line) and  D1 (black dotted  line) profiles.}
\end{figure}

\noindent {\em HD  138813:} A clear, \makebox{very broad ($\Delta$v (FWHM) 
$\sim$18.7 km/s)}  Ca~{\sc ii} K  absorption at -27.5 km/s is  detected, coinciding with  
the ISM velocity (Fig.~\ref{fig:1}), but it  is hardly recognisable  in the
Ca~{\sc ii} H line. The Na {\sc i}  D lines (Fig. 4) show a narrow feature at a
velocity of \makebox{$\sim$ --12 km/s} and EWs of 18\,m\AA ~(D2 line) and
15.4 m\AA ~(D1  line), suggesting that they  are partly optically
thick. Field stars  show  Na {\sc i} D 
absorption features at similar velocities (Appendix A).  Neither Ca {\sc ii} nor 
Na {\sc i} features are detected at the radial velocity of the star.  

\begin{figure}[h] 
\label{HD 1388313_NaI}
\centering
\scalebox{0.6}{\includegraphics{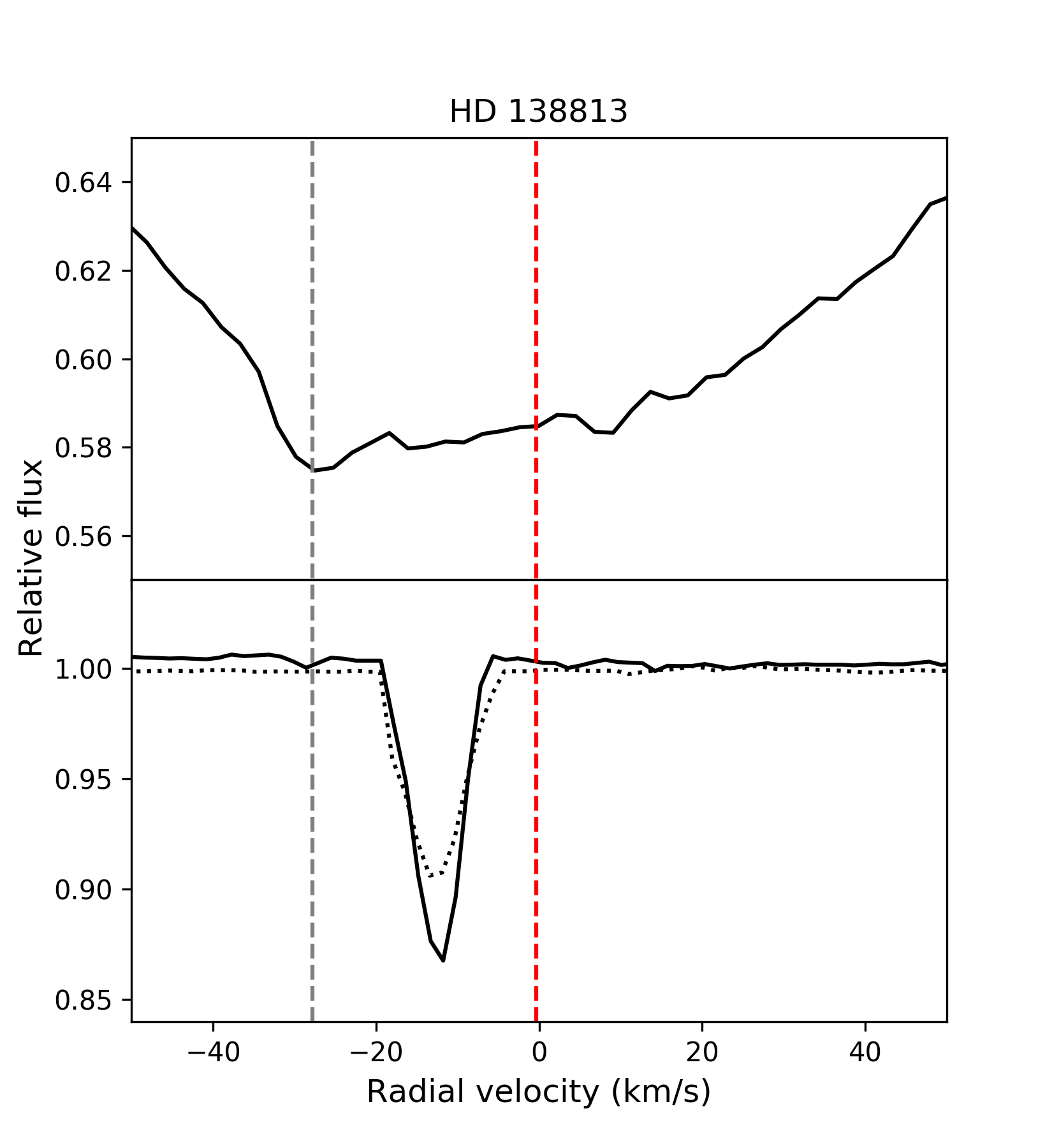}}
\caption{Top panel: Ca   {\sc  ii} K  line   profile  of   HD 138813. 
An absorption feature is detected at the ISM velocity. 
Bottom panel: Non-photospheric Na {\sc i} D2 (black continuous  line) and Na {\sc i} D1 
(black dotted line) absorption features.}
\end{figure}

\noindent {\em HD  146897:} This star presents a narrow  Ca~{\sc ii} K
absorption  at  the  bottom  of  the  photospheric  line  with an  EW of 
$\sim$12.4\,m\AA  \, and  a velocity of  $\sim$ --8.0  km/s,  close to  the
radial    velocity    of    the    star    (Fig.~\ref{fig:1},    Table
\ref{tbl:velocities}).  A second  weaker  feature is  detectable at  a
velocity of  $\sim$ --30.0\,km/s, that is,  the ISM velocity  vector.  No
absorption is recognisable  in the Ca~{\sc ii} H  line.  Narrow strong
Na {\sc  i} D absorptions are  detected with the same  velocity as the
strong Ca~{\sc ii}  K feature (Fig. 5); their EWs  are $\sim$ 118 m\AA
~(D2 line) and $\sim$ 80 m\AA ~(D1 line), suggesting that they are at least
partly  optically thick.  Field  stars around  HD 146897
  show absorption features with similar  velocities in the Ca {\sc ii}
  and/or Na {\sc i} lines (Appendix A).

\begin{figure}[h] 
\label{hd146897_NaI}
\centering
\scalebox{0.6}{\includegraphics{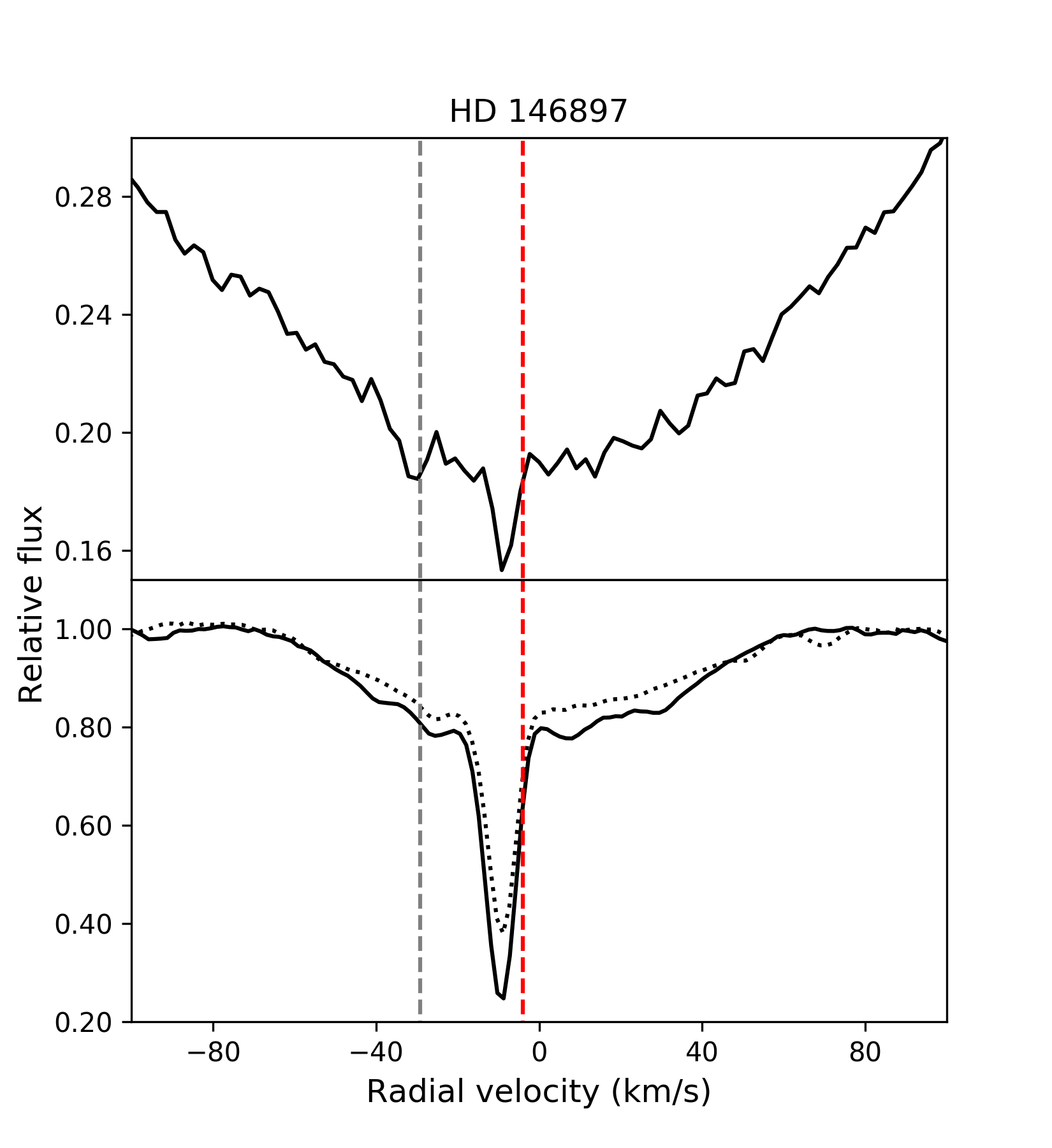}}
\caption{Top panel: Ca   {\sc  ii} K  line   profile  of   HD 146897. 
Bottom panel: Na {\sc i} D2 (black continuous  line) and Na {\sc i} D1 (black
  dotted line).}
\end{figure}

\noindent {\em HD 156623:} Two Ca~{\sc ii} K absorptions at velocities of
$\sim$\,  --15.0 and  $\sim$\,8.1 km/s  and with EWs of  $\sim$ 9.0  m\AA ~and
$\sim$  4.3  m\AA,  respectively,  are detected superimposed  on  the
photospheric line.  Neither feature exactly coincides either with
the  Colorado  model  ISM  velocity vector  or  with the  radial
velocity    of    the    star    (Fig.     \ref{fig:1}    and    Table
\ref{tbl:velocities}). While the strongest, very broad feature remains
practically constant, the weaker one shows variability. As an example,
Fig.  6 shows  three  Ca~{\sc ii}  K  spectra of  HD  156623 taken  at
different epochs  where it  is distinguished that  the $\sim$8.1\,km/s
absorption is clearly  a FEB-like  event. A Ca~{\sc ii}\,H  feature at
the velocity of the strongest  Ca~{\sc ii} K with EW $\sim$\,3.7\,m\AA
~is present in our spectra. Neither CS or ISM absorptions are detected in
the Na {\sc i} D lines.

\begin{figure}[h] 
\label{hd156623_feb}
\centering
\scalebox{0.6}{\includegraphics{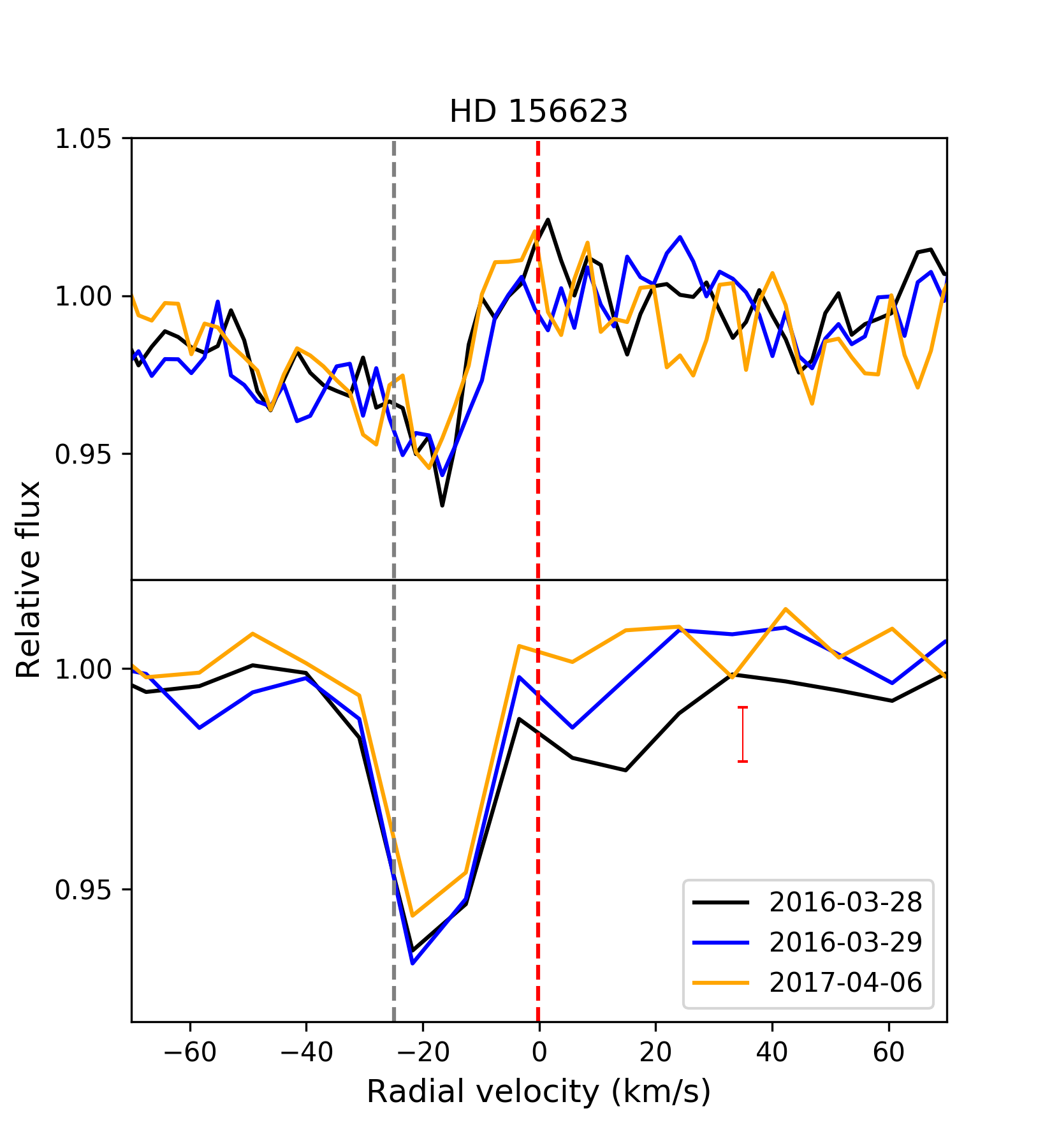}}
\caption{Top  panel:   Individual  non-photospheric  Ca  {\sc   ii}  H
  absorption  features  of  HD  156623. Bottom  panel:Ca  {\sc  ii}  K
  line. The  variable redshifted  absorption at $\sim$  8.1 km/s  is a
  clear  FEB-like  event.  A  3-$\sigma$   error  bar  is  plotted  in
  red. Dates are indicated with different colours in both panels.}
\end{figure}

\noindent  {\em HD  172555:} A  narrow Ca~{\sc  ii} K  absorption with
a velocity of 2.3 km/s and an EW of $\sim$ 9.9 m\AA ~at the  core of the stellar
line is detected (Fig.~\ref{fig:1}  and Table \ref{tbl:velocities}). A
similar absorption  is detected  in the  Ca~{\sc ii}  H line  with an EW(
Ca~{\sc ii} H) of $\sim$ 4.0 m\AA. No feature is detected in the Na\,{\sc
  i}\,D lines.  Similar Ca ~{\sc ii} and Na {\sc i} results were found
by \cite{kiefer14},  although those authors detected  some Ca~{\sc ii}
FEBs not present in any of our spectra.

\noindent  {\em  HD 181296  ($\eta$  Tel):}.  A narrow  absorption  is
observed   in   both   Ca~{\sc   ii}  H\&K   lines   at   a   velocity of
\makebox{$\sim$\,--\,22.8\,km/s} with EWs of $\sim$3.9 ~m\AA ~(K-line) and
$\sim$2.2 ~m\AA ~(H-line).  No feature is detected in the Na {\sc i} D
lines.  The  Ca~{\sc ii} feature  velocity is very different  from the
estimated radial velocity  of the star of -3 km/s.   The line of sight
to  this star  does not  traverse any  identified local  cloud in  the
Colorado model,  but it  is close  ($< 20\degr$) to  some of  them; in
particular to the G and Vel  clouds which have a projected velocity of
--18.51 and --27.6\,km/s, respectively.

\subsection{Stars without narrow absorption components}

Neither  stable nor  variable Ca  {\sc ii}  or Na  {\sc i}  absorption
features, which could be attributed to the ISM or CS medium, have been
detected in  any of  our spectra towards  the  cold-gas-bearing debris-disc stars  HD 21997,  $\eta$  Crv,  HD  121191,  HD 121617,  and  HD
181327. For the sake of completeness,  Fig. 7 shows the observed median
Ca  {\sc  ii} K  profile  of  these  stars  together with  the  Kurucz
synthetic profiles.  A clear emission  due to stellar activity  at the
core of the photospheric line is apparent  in the spectrum of the F6 V
star HD 181327. At  the same time, the observed Ca {\sc  ii} K line of
$\eta$ Crv (HD 109085, F2 V spectral type) is not as deep as predicted
by the  synthetic line.  We have tested the effects of  changing the
metallicity  and  Ca   abundance  in  the  Kurucz   models without finding  a
satisfactory  fit between  the observed  Ca {\sc  ii} K  line and  the
synthetic  one. However, we have  achieved satisfactory  fits
between  Kurucz models  and stars  with similar  properties as  $\eta$
Crv. These  results lead  us to  conclude that   $\eta$ Crv
most likely presents some level of chromospheric activity.

 \begin{figure}[h] 
\label{photospheric}
\centering
\scalebox{0.625}{\includegraphics{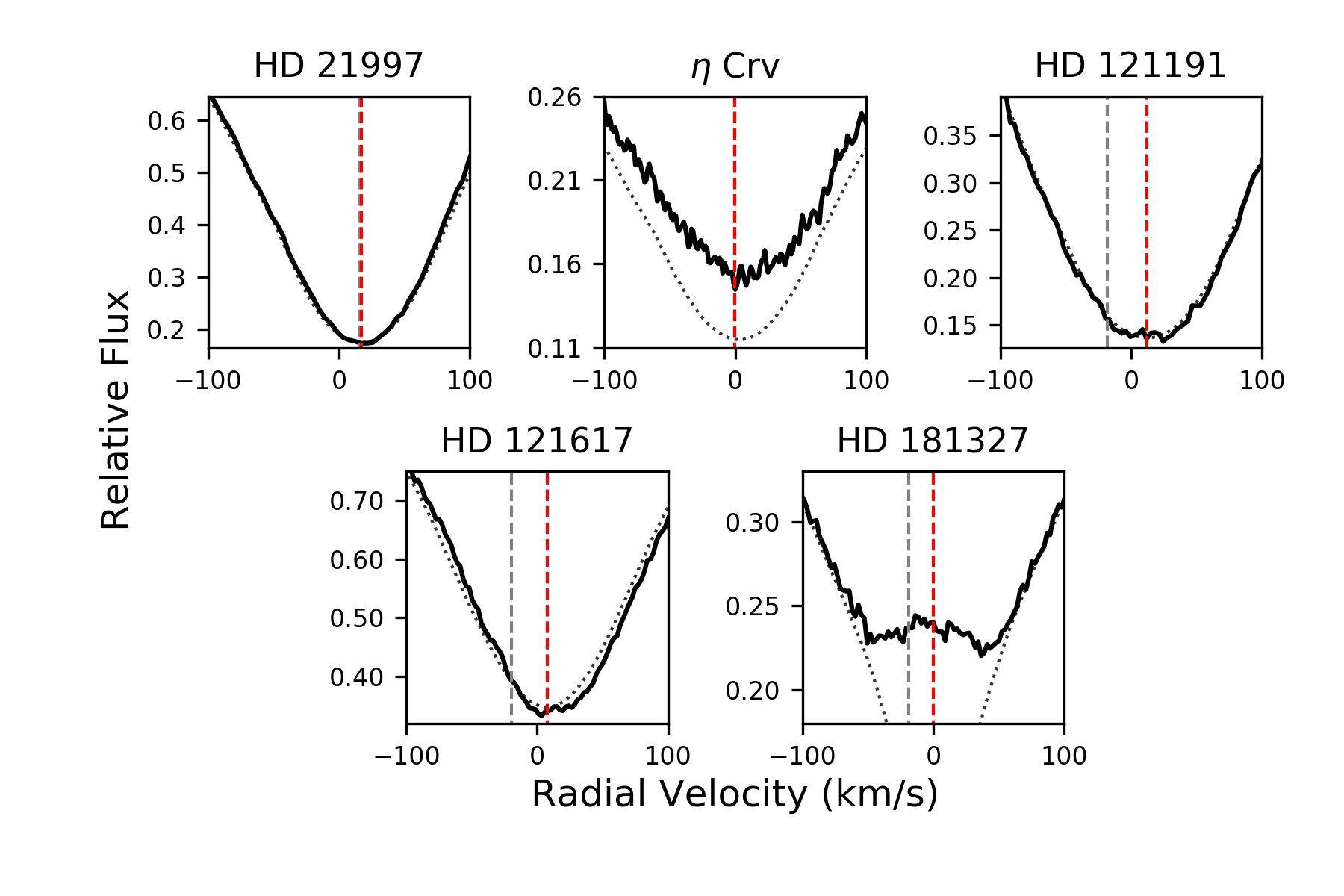}}
\caption{Observed  median  Ca  {\sc  ii}   K  line  profiles  of   cold-gas-bearing debris-disc stars without non-photospheric absorption features.}
\end{figure}

\begin{table*}
\begin{center}
\begin{threeparttable}
\caption {Summary  of the non-photospheric absorption  features of the
  observed cold-gas-bearing debris-disc stars.  Columns  2  to  4:
  Coincidences of the Ca {\sc  ii} velocity components with the radial
  velocity of  the star, Colorado model  ISM clouds along the  line of
  sight, or none of both. Columns 5  to 7: Coincidences of the Na {\sc
    i}  velocity components  with  Ca\,{\sc ii}  features, the  radial
  velocity  of  the  star,  or ISM absorptions.  Column  8  gives  the
  inclination   angle  of   the  debris   disc  associated   with  the
  stars. Empty spaces signify non-detections.}
\begin{tabular}{lccc|ccc|l} 
 \hline 
\hline
\multicolumn{8}{c}{Stars with non-photospheric absorptions} \\
  \hline 
 & \multicolumn{3}{c|}{Ca II} & \multicolumn{3}{c|}{Na I} & Angle (deg.)\tnote{\textit{d}}\\
  \hline 
  Star & V$_{\rm star}$ & V$_{\rm ISM}$ & Neither & V$_{\rm CaII}$ & V$_{\rm star}$ &  V$_{\rm ISM}$& \\
  \hline 
  HD 9672\tnote{\textit{a}}   & Y & Y &   &   &   &   &70-90 (1,2,3)  \\
  \hline
  HD 32297\tnote{\textit{a}}  & Y & Y &   & Y & Y &   &80-88 (4,5)    \\
  \hline
  HD 110058 & Y &   & Y & Y & Y &   & $\sim$90 (6)  \\
  \hline
  HD 131488 & Y & Y  &  & Y & Y &   & 82 (7)         \\
  \hline
  HD 131835                   & Y & Y &  & Y &    & Y\tnote{\textit{c}}  & 75-83 (8,9,14)\\
  \hline
  HD 138813                   &    & Y &   &    &    & Y\tnote{\textit{c}} & 32 (10)        \\
  \hline
  HD 146897                   & Y & Y &   & Y & Y & Y\tnote{\textit{c}}  & 84 (11)        \\
  \hline
  HD 156623\tnote{\textit{a}} &   &   & Y &   &   &   & 34 (10)        \\
  \hline
  HD 172555\tnote{\textit{a}} & Y &   &   &   &   &   & 75 (12)        \\
  \hline
  HD 181298\tnote{\textit{b}} &   &   & Y &   &   &   & $>$70 (13)     \\
  \hline 
 \hline 
\end{tabular}
\begin{tabular}{lccccccl}

 \multicolumn{8}{c}{Stars without non-photospheric absorptions} \\  
  \hline 
 \multicolumn{2}{c}{Star} & HD 21997 & HD 109085 &  HD 121191 & HD 121617 &  \multicolumn{2}{c}{HD 181327} \\
  \multicolumn{2}{c}{Angle (deg.)\tnote{\textit{d}}} & 24 (15) & 35 (16) &  unresol. (7) & 37 (7) &  \multicolumn{2}{c}{30 (17)}   \\
  \hline 
  
\end{tabular}
{\small
\begin{tablenotes}
\item[(\textit{a})] Stars with FEBs. $^{(\textit{b})}$ See text Sec. 5.2. $^{(\textit{c})}$ ISM absorptions from field stars.
$^{(\textit{d})}$ References for inclination angle: (1) \cite{hughes08}; (2)~\cite{moor15b}; (3) \cite{choquet17};
(4) \cite{schneider05}; (5) \cite{asensiotorres16}; (6) \cite{kasper15}; (7) \cite{moor17}; (8) 
\cite{hung15}; (9) \cite{moor15a}; 
(10) Hales et al. (in preparation); (11) \cite{thalmann13}; (12) \cite{smith12};  (13) \cite{smith09};
(14) \cite{feldt17};(15) \cite{moor13}; (16) \cite{marino17}; (17) \cite{marino16} . 
\end{tablenotes}
}
\end{threeparttable}
\end{center}
\label{tbl:summary}
\end{table*}

\section{Discussion}

Before discussing  plausible origins of the  observed non-photospheric
features and their potential correlations  with the properties of the
cold-gas-bearing debris discs and their  host stars, it is useful to  globally
evaluate  the  observational   results  of  the  whole  cold-gas-bearing debris-disc sample, that is,  including the two  systems not
observed by us, $\beta$ Pic and Fomalhaut.
 
\subsection{Global overview of the observational results}
As pointed  out before, 10 out  of the 15 observed  gas-bearing debris-disc stars show  at least one non-photospheric Ca  {\sc ii} absorption
feature. Table 4 summarizes these  observational findings, where we
indicate the stars  with non-photospheric Ca {\sc ii} and  Na {\sc i}
features at the velocities of the star, ISM clouds of the
  Colorado  model, and/or  plausible ISM  absorptions as  suggested by
  their detection in field stars.  We have conservatively assumed that
  the  observed features  have a  stellar  or ISM  counterpart if  the
  velocity difference is less than 5 km/s.  The table also shows those
  stars  with  FEB-like  events.

When considering the whole sample of 17 cold-gas-bearing debris-disc stars, we
find  that 11  of  them  (including $\beta$  Pic)  show  at least  one
non-photospheric Ca {\sc ii} K absorption.  That leaves 6 objects with
no traces  of any  extra absorption  superimposed on  the photospheric
line. Fomalhaut  has been  added to  the observed  stars in  this work
since neither Ca {\sc ii} nor Na {\sc i} extra absorptions in the core
of      the      stellar       lines      have      been      detected
\citep{hobbs86,lagrangehenri90,holweger95}.    Furthermore,  we   have
downloaded UVES  spectra from the  ESO archive confirming the  lack of
a non-photospheric absorption feature in the Fomalhaut spectrum.

\subsection{Origin of the non-photospheric absorption features}

There is some ambiguity concerning the  CS or ISM origin of the stable
Ca {\sc ii}  and Na {\sc i}  absorptions as they can originate in
warm and cold clouds of the local  ISM.  The observed line EWs, their
ratios, and (in most cases) the velocity dispersion of the  Ca {\sc ii} K line are not
different from what is detected in the ISM \citep{redfield02,welsh10},
where different lines of sight with or  without Ca {\sc ii} or Na {\sc
  i}  absorptions are  found, as  well  as a  large range  of Ca  {\sc
  ii}-to-Na {\sc i} ratios \citep{redfield07,welsh10}.  Nonetheless, a
CS origin  has been attributed to  at least one velocity  component in
the  cases   of  49  Ceti,  HD   32297,  HD  110058,  and   HD  172555
\citep[e.g.][]{montgomery12,redfield07,hales17,kiefer14},
respectively.  HD 156623 shows a variable FEB-like event. 
The feature at the stellar radial velocity of HD 131488  is likely CS (Sect. 4.1.1).  
The weak 3$\sigma$ feature, if confirmed, at the radial velocity of HD 131835  
might also be circumstellar. At the same  time,   an  ISM   origin  is  the   
most plausible  case   for  the
  non-photospheric absorptions  in HD  138813 and HD  146897, although
  even in these two stars the Ca {\sc ii} K FWHM dispersion velocities, $\sim$18.7 km/s, 
  are larger than typical ISM values \citep[e.g.][]{redfield02}.

HD 181296 ($\eta$ Tel) merits  particular attention. This star has
a  stable feature blueshifted by  $\sim$  20  km/s  with  respect to  the
stellar     radial     velocity    (Fig.~\ref{fig:1}     and     Table
\ref{tbl:velocities});  its line  of sight  does not  cross any  of the
Colorado model clouds, although it  passes within 20$\degr$ of several
clouds.  $\eta$ Tel  is located  at  a projected  angular distance  of
$\sim$ 7  arcmin from HD 181327,  which corresponds to $\sim$  0.1 pc at
the distance of both stars  (Table \ref{tbl:parameters}). HD 181327 is
another cold-gas debris disc but  without non-photospheric absorption
features  (Table \ref{tbl:velocities}  and  Fig. 7).   Both stars  are
members of  the $\beta$ Pic  moving group, and  form part of  a triple
system \citep{neuhauser11},  sharing distances, radial  velocities, and
proper  motions.   Thus, although  an  ISM  origin cannot  be  totally
excluded from being the $\eta$ Tel Ca {\sc ii} absorption feature, those data
strongly  suggest  that the  feature  is  CS.   We note  that  similar
blueshifted CS  absorptions in UV-excited lines of  C {\sc ii}  and N
{\sc ii} are found around the debris-disc star $\sigma$ Her, which has
a similar $T_{\rm  eff}$ as $\eta$ Tel  \citep{chen03}.  Those authors
suggest that  the blueshifted features  arise in a  radiatively driven
wind, and that the gas could be generated by the sublimation of comets
or collisions between solid bodies.

The previous arguments strongly suggest the  presence of hot Ca {\sc ii}
and Na {\sc i} gas in the  CS environment of at least eight (including
$\beta$ Pic), or  possibly nine (if the feature at  the stellar radial
velocity of HD 131835 is confirmed) of the cold-gas-bearing debris-disc stars.
Furthermore, the  presence/absence of Ca  {\sc ii} and Na  {\sc i}
suggests that there  are differences in the CS  gas composition around
the debris-disc stars. 
We note that  \cite{kiefer15} suggest the presence of  two families of
exocomets in $\beta$ Pic, one  of them strongly depleted in volatiles.
Assuming, therefore,  the plausible circumstellar-hot-gas scenario as
suggested by some of the  non-photospheric features, the permanence of
such gas in  the surroundings of the stars would  require a continuous
replenishment  as it  would be  subject  to a  strong radiation  field
overcoming  the gravitational  force \citep{beust89}.   These authors,
and  \cite{lagrange98}, \cite{liseau03},  and \cite{fernandez06}  have
estimated the ratio $\beta = F_{\rm  rad}/F_{\rm grav}$ in the case of
$\beta$ Pic for  lines of many different  atoms/species, including the
Ca {\sc ii} H\&K  and Na {\sc i} lines, and shown  that the atomic gas
would disappear unless a braking mechanism is at work. This would also
be the  scenario of the stars  in this work as  their radiation fields
are similar to that of $\beta$ Pic.
 
Thus, a braking mechanism should be  at work to explain the hot, inner
($<$ 1  au) CS gas around  these gas-bearing debris-disc  stars.  Such
a mechanism  could   be  the  one  suggested   by  \cite{fernandez06}  -
acceleration  of  an  ionic  fluid  produced  by  an  enhanced  carbon
abundance.   In this  sense, it  would  be interesting  to search  for
carbon  lines  around these  stars, for example  in  the UV  spectral  range
\citep{roberge06,roberge14}.

\subsection{Correlation of the non-photospheric features with the inclination of the 
debris discs}

We have checked  if any observational  property of the
stars or of the gas/dust components in their debris discs correlate in
some way  with the  presence/absence and  strength of  the \makebox{Ca
  {\sc  ii}/Na  {\sc  i}}  absorptions  at the  core  of  the  stellar
lines. We note that  CS features  are only detected in the A-type stars 
of the sample and in none of  the three F-type stars, but 
we  cannot identify  any trend  with other properties  like 
rotational velocity or age.

Concerning the debris disc properties, the  only trend we find is that
non-photospheric CS absorptions  are predominantly detected  when the
discs  are oriented  close to  edge-on. This is the case in  at least  7 
  (possibly   8 if HD 131835 is included)  out   of  9   objects.  On the 
  contrary, discs seen close to face-on (8 objects) do not show 
  non-photospheric absorptions, with the exception of  HD 156623 that 
  shows evidence of hot gas as suggested by the FEB-like absorption
(Table 4). $\beta$ Pic and Fomalhaut are included in those figures, as the $\beta$
Pic disc is seen edge-on, $\sim$ 90$\degr$ \citep{smith84}, but not the
Fomalhaut disc, $\sim$ 65$\degr$ \citep{acke12}.  Thus, at
  least $\sim$ 78\% (possibly 88\%) of edge-on cold-gas-bearing debris-disc stars show evidence of hot CS  gas, while only $\sim$ 12 \% of
  the face-on cold-gas-bearing debris-disc stars show such evidence. While
this result is not surprising as such, it suggests that the non-detection  is  most likely
due to geometrical effects, that is, the disc inclination with respect to
the line of sight,  but not to the absence of hot  gas in those discs.
 In  the case of HD  156623,  the detection of the 
FEB-like Ca {\sc ii} event, in  addition  to  the  very broad  stable
absorption (Table 3), might indicate a significant  scale height of the hot gas, 
 since such gas would  be located very close to the  star.   Based on the CO
gas  detection  in  $\beta$ Pic,  \cite{kral16,kral17}  have  recently
developed  a model  predicting the  detection  with ALMA  of cold  gas
around  debris-disc  stars  assuming  that the  gas  arises from  icy,
volatile-rich planetesimals.  

From   the  observational
perspective presented in this work,  our results strongly suggest that
debris discs with  central absorptions indicative of hot  CS gas would
be good targets to search for a cold-gas counterpart with ALMA.

With respect to  other  debris-disc properties, all but  HD 121191 have
been  resolved either  in their  dust thermal  emission, gas  emission,
and/or scattered light \citep[e.g.][see also the Catalogue of Resolved
  Debris
  Disc\footnote{http://www.astro.uni-jena.de/index.php/theory/catalog-of-resolved-debris-disks.html}
  and   references   therein]{asensiotorres16,schneider05,schneider14,
  feldt17,choquet17,moor17}.   Concerning the  cold  gas, neither  the
detected species nor  the flux/luminosity of the lines  show any trend
with the optical absorption lines - see e.g. \cite{moor17} for CO line
luminosities.   With regard  to the  dust, all  systems but  HD 172555
present  two-temperature SEDs. \cite{kennedy14}  and \cite{geiler17}
have analysed  large samples  of two-temperature debris-disc systems,
and  have  shown  that  most  of them  likely  have  a  two-belt  disc
structure, similar to the  solar system's zodiacal and  Kuiper belts. There
are, however, a  small number of exceptions. In  particular, HD 181327
seems  to be  surrounded by  an  extended disc  with a  range of  dust
temperatures and  properties \citep{lebreton12,kennedy14,ballering17}.
Several works  postulate an exo-asteroid belt  and/or cometary sources
as     the      possible     origin      of     the      warm     disc
\citep{kennedy14,ballering17,geiler17}.   In any  case, although  both
dust  belts and hot  gas likely share a similar, comet-related
origin, we  cannot identify any  trend of  the dust-disc  properties -
fractional luminosities,  radii, and temperatures  - with the  presence or
strength of  the hot-gas absorptions.  It is interesting to  note that
the prototype  FEB host $\beta\,Pic$  also harbours species of very hot  dust  in addition  to hot
gas \citep{def12}.  HD\,172555
also harbors both hot dust and gas \citep{ert14, ert16}. The origin of
this hot dust is still unclear.  Although the samples of stars checked
for hot dust and those for hot gas have little overlap so far, and thus
a correlation cannot be seen, it seems conceivable that both phenomena
might be related.

\section{Conclusions}

In this work we present  high-resolution optical spectra of the
currently known cold-gas-bearing debris-disc systems. The observations
show Ca  {\sc ii} and Na  {\sc i} non-photospheric absorptions  in  at least 
$\sim$ 78 \%, and possibly 88\%, of the 
discs  seen edge-on, and in only $\sim$ 12\% of the face-on discs.    Those  
absorptions are  usually interpreted  as signposts of  hot gas  released by  
exocomets or  small grains  in the close  vicinity  of the  central  stars.  
Thus, these detection rates suggest  that  the non-detections  in  some  
discs  seen  close to  face-on  are most likely due  to
geometrical effects, that is, the inclination of the CS disc. At the same
time, hot  gas is only detected in A-type debris-disc systems, but it  does not correlate 
either with any other  stellar or disc properties.

The  optical, infrared  and  submillimetre  observations suggest  that
planetesimals/cometary bodies  often populate  gas-bearing debris-disc  systems,   from  a   few  stellar  radii   up  to   hundreds  of
au. Collisions among  those large bodies, as well as  among small dust
grains, and their evaporation are  the agents, possibly accompanied by
planets, as in the case of $\beta$ Pic, structuring the CS environments
and planetary  systems of  main-sequence stars.  \cite{choquet17} have
recently outlined a scenario to interpret  the gas and dust structure of the 49
Ceti system; the authors suggest physical mechanisms for the production of
gas at  different distances from the central  star.  That scenario
is likely at  least roughly applicable to the systems  studied in this
work.

\begin{acknowledgements}
Based on  observations made with  the Mercator Telescope,  operated on
the  island of  La Palma  by the  Flemmish Community,  and the  Nordic
Optical Telescope, operated by the Nordic Optical Telescope Scientific
Association, at the Spanish Observatorio del Roque de los Muchachos of
the Instituto de  Astrofísica de Canarias. Based  on observations made
with  ESO Telescopes  at  the La  Silla  Observatory under  programmes
099.A-9004(A) and 099.A-9029(A). Also  based on observations made with
the  TIGRE  telescope  funded  and operated  by  the  universities  of
Hamburg, Guanajuato and  Li\`ege. Partially based on data obtained from the ESO Science Archive Facility. Paula Sarkis kindly  provided observations of
HD 121191. We thank A. Hales  for sharing with us the disc inclination
of HD 138813 and HD 156623 before publication. H.C., C.E., A.G., G.M.,
B.M.,   I.R.,  and   E.V.   are   supported  by   Spanish  grant   AYA
2014-55840-P.  H.C.   acknowledges  funding  from  the   ESA  Research
Fellowship Programme. D.P.I. thanks Iv\'an Lacerna for his help during
the observations in La Silla in  October 2015.  This work has made use
of  data from  the  European  Space Agency  (ESA)  mission {\it  Gaia}
(\url{https://www.cosmos.esa.int/gaia}), processed  by the  {\it Gaia}
Data      Processing      and     Analysis      Consortium      (DPAC,
\url{https://www.cosmos.esa.int/web/gaia/dpac/consortium}).    Funding
for the DPAC has been provided by national institutions, in particular
the  institutions   participating  in  the  {\it   Gaia}  Multilateral
Agreement.  The  work of A.M. was  supported by the Momentum  grant of
the MTA CSFK  Lend\"ulet Disk Research Group and  Hungarian OTKA grant
K125015.
We thank the anonymous referee for his/her comments and suggestions, which helped to improve the paper.
\end{acknowledgements}

%
\bibliographystyle{aa} 
\bibliography{febs} 

\begin{appendix}

\section{Field stars}  

This  section  provides  information  on  non-photospheric  absorption
features of field stars around HD 131488, HD 131835, HD 138813, and HD
146897. These four stars were selected due to the ambiguous origin of their non-photospheric absorptions. The data are based on a search of stars in a field of 5$\degr$
in radius  around each of the four stars, and  parallaxes between 6 and  10 mas, so
that the field  stars are located at comparable distances  of the cold-gas-bearing debris-disc stars. We recall  that they are members of the
Scorpius-Centaurus OB association. Spectra of the field stars have
been retrieved from the ESO archive of
FEROS\footnote{\tt http://archive.eso.org/wdb/wdb/adp/phase3\_spectral/ form?collection\_name=FEROS},
HARPS\footnote{\tt http://archive.eso.org/wdb/wdb/eso/repro/form} 
and UVES\footnote{\tt http://archive.eso.org/wdb/wdb/eso/uves/form}, as well as our own 
observations in the case of HD 146897. Field stars around each debris-disc
star, their angular distances, stellar radial velocities, and  velocities
of non-photospheric features are given in Tables A.1 to A.4.

\noindent  {\bf \em HD  131488}.  The field  stars  HD  132058, HD  133937, and  HD 135454  
have features  at $\sim$ --20 km/s (Table A.1), relatively close in velocity to the 
velocity vectors
of the G and Mic clouds in  the Colorado model.  HD 128207, HD 131120,
HD 132094,  and HD 133880 have  a common feature at  $\sim$ --13 km/s,
suggesting it is an interstellar absorption not traced in the Colorado
model. The star HD 132200 presents  two features at --6.9 km/s and 4.6
km/s, which are close to the ones observed in  HD 131488 at --3.2 km/s 
and 4.0 km/s.  The HD  132200 feature  at -6.9  km/s remains
practically constant. This absorption, and the -3.2 km/s of our  debris-disc star, are close  to the mean velocity of $\sim$  -- 6.6 km/s that
characterize  the  approaching  face  of  the  bubble  around  the  OB
association \citep{poppel10}.  Thus, they could  have an ISM  origin. At
the same  time, the  HD 132200 4.6  km/s feature  varies significantly
(Fig. A.1),  which suggests it  is related  to the star.   This fact
also suggests that  the 4.0 km/s absorption in HD 131488, which  coincides with the
stellar radial velocity, is likely circumstellar.

\begin{table}[h]                                                              
\caption{HD 131488: field stars}                                           
\begin{tabular}{lrrr}                                                      
\hline                                                                     
\hline                                                                     
Star             & r       & v$_{rad}$ & v$_{CaII K}$  \\                   
                   & (deg.)& (km/s)     & (km/s)           \\              
\hline                                                                    
HD 131488 &         & 5.8 & --3.2/4.0\\                                    
\hline                                                                     
HD 128207 & 3.60 & 2.0 & --13.0 \\                                         
HD 131120  & 3.35 & 6.1 & --16.0 \\                                        
HD 132058  & 2.11 & 0.2 & --19.1 \\                                        
HD 132094 & 3.81 & 0.9 & --11.4 \\                                        
HD 132200  & 1.24 & 8.0 & --6.9/4.6 \\                                     
HD 133880  & 2.53 & 2.8 & --13.0 \\                                        
HD 133937  & 3.06 & 2.0 &  --21.4 \\                                       
HD 135454  & 4.20 & 1.4 & --24.4 \\                                        
\hline                                                                     
\end{tabular}                                                               
\end{table}   
                                                           
\noindent{\bf \em HD 131835}.  All field stars, as well as HD 131835, share the
{\makebox{$\sim$ --13.0}}  km/s feature  mentioned above  (Table A.2),
which likely  has an ISM origin.   One of the field  stars, HD 132955,
presents  a second,  strong  feature at  {\makebox{$\sim$ 3.8  km/s}}.  
Our star,  HD 131835, also  has a very weak  (requiring confirmation)  2.2 km/s  feature at  the stellar
radial velocity. If  real, its CS or ISM origin is ambiguous.  None
of the field stars have a feature close  to the one at -5.9 km/s of HD
131835.   This absorption  might correspond  to the  one at  --6.6 km/s
which characterizes the mentioned bubble of the OB association.

\begin{table}[h]                                                             
\caption{HD 131835: field stars.}                                          
\begin{tabular}{lrrr}                                                     
\hline                                                                    
\hline                                                                     
Star             & r       & v$_{rad}$ & v$_{CaII K}$  \\                  
                   & (deg.)& (km/s)     & (km/s)           \\             
\hline                                                                     
HD 131835 &           & 2.6    & -13.8/-5.9/2.2 \\                         
\hline                                                                     
HD 131120  & 2.26 & 6.1 & --16.0 \\                                       
HD 132094 & 1.69 & 0.9 & --11.4 \\                                         
HD 132955 & 3.30 & 9.5 & --13.7/3.8 \\                                     
\hline                                                                    
\end{tabular}                                                             
\end{table}
                                                               
\noindent{\bf \em HD 138813}.  Like HD  138813, all field stars show a
non-photospheric absorption Na {\sc i} D2  at $\sim$ -- 13 km/s (Table
A.3), strongly suggesting an ISM origin.
                                                                          
\begin{table}[h]                                                              
\caption{HD 138813: field stars.}                                            
\begin{tabular}{lrrr}                                                      
\hline                                                                    
\hline                                                                    
Star             & r       & v$_{rad}$ & v$_{NaI D2}$  \\                   
                   & (deg.)& (km/s)     & (km/s)           \\              
\hline
HD 138813 &             &--0.4    & -12.0  \\                              
\hline                                                                     
HD 141637 &  3.54 & --3.0 &  --13.3/-8.7\\    
HD 136246 &  4.16 & -2.7   &  --16.0\\ 
HD  142301 &  4.40 & -8.7 &  --14.3\\                                     
HD 142184  & 4.58 & -9.2  &  --13.3 \\                                    
HD 142250  & 4.59 & --1.3 & --12.8 \\                                     
\hline                                                                    
\end{tabular}                                                              
\end{table}                                                               

\noindent{\bf \em HD 146897.}        
Absorption features at velocities close to the ones of HD 146897 are detected 
in the Ca {\sc ii} and/or Na {\sc i} lines of the field stars (Table A.4). This suggests 
an ISM origin for the HD 146897 non-photospheric absorptions.

\begin{table}[h]                                            
\caption{HD 146897: field stars.}                                          
\begin{tabular}{lrrrr}                                                    
\hline                                                                    
\hline                                                                    
Star             & r       & v$_{rad}$ & v$_{CaII K}$ & v$_{NaI D2}$  \\    
                   & (deg.)& (km/s)      & (km/s)            & (km/s)\\   
\hline                                                                     
HD 146897 &          &--4.0          & --30.0,--8.0    & -8.8       \\     
\hline                                                                     
HD 144587 &  3.76 & 0.0            & --                   & --22.5,--9.2   \\ 
HD 145554 &  2.47 & --6.1         &  --25.9,--9.2    &  --25.5,--10.2 \\   
HD 145631 &  2.47 & --5.6         &  --27.5,--9.2    & --25.5,--10.2\\    
HD 145964 & 1.20  & --7.8         &  --25.2,-9.2     &  --22.5,--9.2\\     
HD 147137 &  1.23 & --0.8         &  --9.2              &  --8.7\\         
HD 147932 &  2.45 &                  & --8.4               &  --9.5\\     
\hline                                                                     
\end{tabular} 
\end{table}

\begin{figure}[h!]
\centering                                                                                                                                                                         
\scalebox{0.5}{\includegraphics{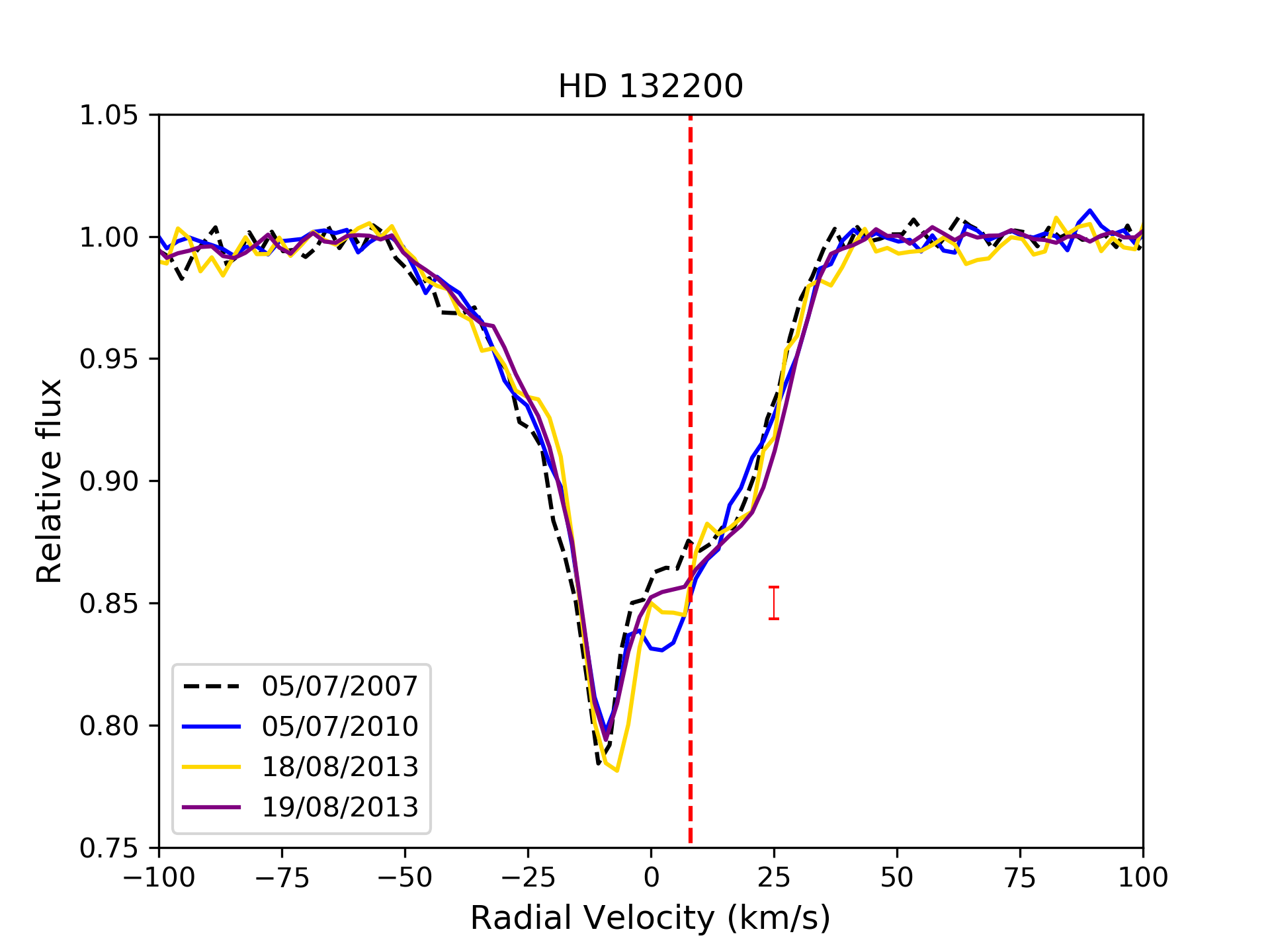}}
\caption{Ca {\sc ii} K line profiles of HD 132200. Solid lines are for FEROS spectra and the dashed line is for a HARPS spectrum. Observing dates are indicated. Vertical red dashed line indicates the radial velocity of the star. A 3-$\sigma$ error bar is plotted in red.}
\label{fig:HD132200} 
\end{figure}

\section{Observing log}

\begin{table}[!h]                                                                                          
\caption{Observing Log: Stars, Dates (Universal Time) and Instruments.}                           
\label{table:observing}                                                                                       
\centering                                                                                              
\begin{tabular}{lllllll}   
\hline                                                                                                   
Star & Date (UT)& Instrument & Date (UT) & Instrument & Date (UT) & Instrument\\                      
\hline 
HD 9672 &20150904T0107 & HERMES       & 20150904T0143 & HERMES       & 20150904T0318 & HERMES   \\   
(49 Cet)&20150904T0539 & HERMES       & 20150905T0117 & HERMES       & 20150905T0133 & HERMES   \\   
        &20150905T0335 & HERMES       & 20150905T0523 & HERMES       & 20150906T0126 & HERMES   \\   
        &20150906T0221 & HERMES       & 20150906T0237 & HERMES       & 20150906T0529 & HERMES   \\   
        &20150907T0113 & HERMES       & 20150907T0129 & HERMES       & 20150907T0414 & HERMES   \\   
        &20150907T0430 & HERMES       & 20151220T2047 & HERMES       & 20151220T2108 & HERMES   \\   
        &20151222T2122 & HERMES       & 20151222T2148 & HERMES       & 20151222T2213 & HERMES   \\   
        &20151223T2059 & HERMES       & 20151223T2230 & HERMES       & 20160126T2000 & FIES     \\   
        &20160126T2044 & FIES         & 20160127T1940 & FIES         & 20160127T1949 & FIES     \\   
        &20160127T1954 & HERMES       & 20160127T2006 & HERMES       & 20160128T1934 & HERMES   \\   
        &20160128T1946 & HERMES       & 20160130T2004 & HERMES       & 20160130T2017 & HERMES   \\   
        &20160714T0505 & HERMES       & 20160715T0457 & HERMES       & 20160717T0510 & FIES     \\   
        &20160718T0502 & FIES         & 20160719T0453 & FIES         &               &          \\   
\hline                                                                                               
HD 21997 &20150904T0503 & HERMES       & 20150905T0422 & HERMES       & 20150905T0443 & HERMES  \\  
         &20150906T0448 & HERMES       & 20150906T0508 & HERMES       & 20150907T0451 & HERMES  \\  
         &20150907T0512 & HERMES       & 20150919T0453 & TIGRE        & 20150920T0251 & TIGRE   \\  
         &20150924T0340 & TIGRE        & 20150925T0311 & TIGRE        & 20150926T0440 & TIGRE   \\  
         &20151023T0225 & FEROS        & 20151024T0526 & FEROS        & 20151023T0611 & FEROS   \\  
         &20151110T2302 & TIGRE        & 20151111T2230 & TIGRE        & 20151114T0058 & TIGRE   \\  
         &20151114T2216 & TIGRE        & 20151220T2220 & HERMES       & 20151220T2251 & HERMES  \\  
         &20151223T2304 & HERMES       & 20160127T2030 & HERMES       & 20160128T2011 & HERMES  \\  
         &20160130T2050 & HERMES       &               &              &               &         \\  
\hline                                                                                                   
HD 32297 &20150904T0405 & HERMES       & 20150904T0436 & HERMES       & 20151221T0106 & HERMES  \\     
         &20151221T0137 & HERMES       & 20151221T0208 & HERMES       & 20151223T0050 & HERMES  \\     
         &20151223T0121 & HERMES       & 20151224T0159 & HERMES       & 20151224T0230 & HERMES  \\     
         &20160126T2148 & FIES         & 20160126T2220 & FIES         & 20160127T2037 & FIES    \\     
         &20160127T2108 & FIES         & 20160127T2231 & HERMES       & 20160127T2301 & HERMES  \\     
         &20160128T2204 & HERMES       & 20160128T2235 & HERMES       & 20160130T2231 & HERMES  \\     
         &20160130T2302 & HERMES       &               &              &               &         \\     
\hline                                                                                                   
HD 109085 &20151221T0642 & HERMES      & 20160129T0528 & HERMES       & 20160131T0305 & HERMES  \\    
($\eta$ Crv) &              &             &               &              &               &         \\    
\hline                                                                                                
HD 110058  &20170402T0440& FEROS        & 20170403T0414& FEROS         & 20170404T0402 & FEROS  \\   
(HIP 61782)&20170405T0329& FEROS        & 20170406T0331& FEROS         & 20170407T0418 & FEROS  \\   
           &20170408T0251& FEROS        & 20170409T0410& FEROS         &               &        \\   
\hline                                                                                                
HD 121191 & 20170601T0121 & FEROS       &              &               &               &        \\   
\hline                                                                                                
HD 121617 & 20170409T0238 & FEROS       & 20170409T0611 & FEROS        &               &        \\  
\hline                                                                                                
HD 131488 & 20170409T0219 & FEROS       & 20170409T0551 & FEROS        &               &        \\  
\hline                                                                                                
HD 131835 &20160326T0408 & FEROS       & 20160326T0755 & FEROS       & 20160327T0322 & FEROS    \\   
(HIP 73145)&20160327T0544& FEROS       & 20160327T0754 & FEROS       & 20160328T0315 & FEROS    \\  
          &20160328T0455 & FEROS       & 20160328T0808 & FEROS       & 20160329T0304 & FEROS    \\     
          &20160329T0429 & FEROS       & 20160329T0836 & FEROS       & 20170403T0519 & FEROS    \\    
          &20170407T0456 & FEROS       & 20170408T0322 & FEROS       &               &          \\    
\hline                                                                                                   
HD 138813  &20160304T0412 & HERMES     & 20160304T0443 & HERMES       & 20160305T0407 & HERMES \\        
(HIP 76310)&20160305T0438 & HERMES     & 20160305T0509 & HERMES       & 20160306T0418 & HERMES \\        
           &20160306T0449 & HERMES     & 20160326T0424 & FEROS        & 20160326T0817 & FEROS  \\        
           &20160327T0336 & FEROS      & 20160327T0626 & FEROS        & 20160327T0828 & FEROS  \\        
           &20160328T0421 & FEROS      & 20160328T0833 & FEROS        & 20160329T0506 & FEROS  \\        
           &20160329T0901 & FEROS      & 20160711T2142 & HERMES       & 20160712T2233 & HERMES \\        
           &20160712T2304 & HERMES     & 20160713T2148 & HERMES       & 20160713T2219 & HERMES \\        
           &20160714T2122 & HERMES     & 20160716T2128 & FIES         & 20160717T2148 & FIES   \\        
           &20160718T2202 & FIES       & 20160719T2230 & FIES         & 20170309T0518 & HERMES \\        
           &20170310T0610 & HERMES     & 20170310T0639 & HERMES       & 20170314T0452 & HERMES \\        
           &20170314T0523 & HERMES     & 20170314T0554 & HERMES       & 20170403T0452 & FEROS  \\        
           &20170404T0421 & HERMES     & 20170405T0505 & FEROS        & 20170406T0443 & FEROS  \\        
           &20170408T0354 & FEROS      &               &              &               &        \\        
\hline                                                                                                   
HD 146897  &20160712T2115 & HERMES       & 20160712T2146 & HERMES        &20170405T0702 & FEROS       \\ 
(HIP 79977)&20170405T0733 & FEROS        & 20170407T0726 & FEROS         &20170407T0756 & FEROS       \\ 
\hline                                                                                                   
\end{tabular}                                                                                            
\end{table}                                                                                             
\setcounter{table}{0}                                                                                    
\begin{table*} [!h]                                                                                           
\caption{Log of observations:continuation}                                                               
\label{table:observing}                                                                                       
\centering                                                                                               
\begin{tabular}{lllllll}                                                                                 
\hline                                                                                                   
Star & Date & Telescope & Date & Telescope & Date & Telescope\\                                          
\hline 
HD 156623  &20151022T0514 & FEROS  & 20160328T0520 & FEROS & 20170405T0839 & FEROS\\                     
(HIP 84881)&20160326T0438 & FEROS  & 20160328T0858 & FEROS & 20170406T0619 & FEROS\\                     
           &20160326T0838 & FEROS  & 20160329T0655 & FEROS & 20170408T0717 & FEROS\\                     
           &20160327T0441 & FEROS  & 20160329T0925 & FEROS &               &      \\                     
           &20160327T0902 & FEROS  & 20170402T0931 & FEROS &               &      \\                     
\hline                                                                                                    
HD 172555 &20151022T0356 & FEROS  & 20151023T0238 & FEROS  & 20160326T0534 & FEROS  \\                   
          &20160326T0849 & FEROS  & 20160327T0715 & FEROS  & 20160327T0958 & FEROS  \\                   
          &20160328T0638 & FEROS  & 20160328T0943 & FEROS  & 20160329T0706 & FEROS  \\                   
          &20160329T0934 & FEROS  & 20170402T0629 & FEROS  & 20170403T0721 & FEROS  \\                   
          &20170404T0707 & FEROS  & 20170405T0657 & FEROS  & 20170406T0703 & FEROS  \\                   
          &20170407T0656 & FEROS  & 20170408T0647 & FEROS  &               &        \\                   
\hline                                                                                                   
HD 181296 &  20151022T0357 & FEROS  & 20151023T0041 & FEROS  & 20151024T0309 & FEROS \\                  
($\eta$ Tel) &  20160326T0620 & FEROS  & 20160327T0719 & FEROS  & 20160327T1004 & FEROS \\                  
          &  20160328T0644 & FEROS  & 20160328T0949 & FEROS  & 20160329T0712 & FEROS \\                  
          &  20160329T0942 & FEROS  & 20170403T0726 & FEROS  & 20170404T0739 & FEROS \\                  
          &  20170405T0815 & FEROS  & 20170406T0802 & FEROS  & 20170407T0840 & FEROS \\                  
          &  20170408T0652 & FEROS  &               &        &               &       \\                  
\hline                                                                                                   
HD 181327 & 20170403T0731 & FEROS  & 20170404T0713 & FEROS  & 20170406T0806 & FEROS  \\                  
          & 20170407T0845 & FEROS  & 20170408T0657 & FEROS  &               &        \\                  
\hline                                                                                                   
\end{tabular}                                                                                            
\end{table*}                                                                                             
\end{appendix}
\end{document}